\documentclass[manuscript]{aastex}
\usepackage{natbib,graphicx, amsmath}

\begin{document}

\title{Direct Calculation of the Turbulent Dissipation Efficiency in Anelastic
Convection}

\author{Kaloyan Penev}
\affil{60 Garden St., M.S. 10, Cambridge, MA 02138}
\author{Joseph Barranco}
\affil{1600 Holloway Avenue, San Francisco, CA 94132-4163}
\author{Dimitar Sasselov}
\affil{60 Garden St., M.S. 16, Cambridge, MA 02138}

\begin{abstract}
The current understanding of the turbulent dissipation in stellar convective
zones is based on the assumption that the turbulence follows Kolmogorov scaling.
This assumption is valid for some cases in which the time frequency of the
external shear is high (e.g. solar p-modes). However, for many cases of
astrophysical interest (e.g. binary orbits, stellar pulsations etc.) the
timescales of interest lie outside the regime of applicability of Kolmogorov
scaling.
We present direct calculations of the dissipation efficiency of the turbulent
convective flow in this regime, using simulations of anelastic convection with
external forcing. We show that the effects of the turbulent flow are well
represented by an effective viscosity coefficient and we provide the values of
the effective viscosity as a function of the perturbation frequency and compare
our results to the perturbative method for finding the effective viscosity 
of \citet{Penev_Sasselov_Robinson_Demarque_08} that can be applied to actual
simulations of the surface convective zones of stars.
\end{abstract}

\section{Introduction}
\label{sec: intro}
For stars with surface convection, turbulent dissipation in the convective
zone is believed to be
the dominant mechanism responsible for the conversion of mechanical energy of
tides, stellar oscillations and stellar pulsations to heat. Thus, this is the
mechanism believed to determine the rates of tidal synchronization and
circularization \citep{Zahn_66, Zahn_89}, the amplitudes of stellar $p$ 
modes \citep{Goldreich_Keeley_77, Goldreich_Kumar_88,
Goldreich_Kumar_Murray_94}, and the instability of stars to pulsations
\citep{Gonczi_82}.

The simplest approach to estimating the dissipation efficiency of the turbulent
convection is to assume that the turbulence is homogeneous and isotropic and its
interaction with external shear is only local. In that case, we can define an
effective viscosity coefficient which will capture the effects due to the
turbulent flow. The question then is how to find the value of this coefficient. 

Currently two prescriptions, based on the assumption of Kolmogorov cascade,
exist for estimating this coefficient as a function
of the time period of the external shear ($T$).

\citet{Zahn_66, Zahn_89} proposes that, the effective viscosity should scale
linearly with the fraction of a turn eddies manage to complete in half a
perturbation period:
\begin{equation}
	\nu = \nu_{max} \min\left[ \left(T\over2\tau\right),
	1\right].
	\label{eq: Zahn viscosity}
\end{equation}

On the other hand, \citet{Goldreich_Nicholson_77} and 
\citet{Goldreich_Keeley_77}, argue that eddies with turnover times
bigger than $T/2\pi$ will not contribute
to the dissipation. Then Kolmogorov scaling predicts:
\begin{equation}
	\nu = \nu_{max} \min\left[\left( T\over 2\pi\tau\right)^2,
	1\right]
\end{equation}

Zahn's more efficient dissipation seems to be in better agreement with
observations of tidal circularization times for binaries containing a giant
star \citep{Verbunt_Phinney_95}, the location of the red edge of the Cepheid
instability strip \citep{Gonczi_82}, and even this more efficient prescription
might be insufficient to explain the main sequence circularization of binary
stars in clusters \citep{Meibom_Mathieu_05}.

However, \citet{Goldreich_Keeley_77}, \citet{Goldreich_Kumar_88} and
\citet{Goldreich_Kumar_Murray_94}, successfully used the less efficient
dissipation to develop a theory for the damping of the solar $p$-modes. In this
case the more effective dissipation would require dramatic changes in the
excitation mechanism in order to explain the observed amplitudes.

Finally, \citet{Goodman_Oh_97} calculated a lowest order expansion of the
effective
viscosity, which when applied to Kolmogorov turbulence predicts a result closer
to the less efficient Goldreich \& Nicholson viscosity. While this gives a
firmer theoretical foundation for the quadratic prescription it does not
help with the observational problem of insufficient dissipation in the case of
tides and stellar pulsations. 

A possible resolution of this problem is suggested by the fact that the
successful applications of the two prescriptions correspond to very different
perturbation periods. The \citet{Zahn_66, Zahn_89} scaling seems to work well
for periods of order days, and the Goldreich and collaborators quadratic scaling
seems to apply for periods of order minutes. This distinction is important
because, in stars with surface convection, Kolmogorov scaling predicts that the
eddies with turnover times of several minutes would have typical sizes that are
very small compared to the local pressure scale height and any other external
length scales. On the other hand turnover times of days correspond to eddies
with typical sizes comparable or larger than the local pressure scale height. In
this case Kolmogorov scaling is not expected to apply. 

The flow seen in 2D and 3D
simulations of stellar convection 
is very different from a Kolmogorov cascade \citep{Sofia_Chan_84,
Stein_Nordlund_89, Malagoli_Cattaneo_Brummell_90}. There are two important
distinctions. The first is that the velocity power spectrum is much flatter in
the simulations than Kolmogorov, and so one expects to find a slower loss of
dissipation efficiency with increased frequency of the external shear, assuming
that the dissipation is dominated by the resonant eddies, as long
as the external shear has a period that corresponds to eddy turnover times
too long to fall in the inertial subrange of Kolmogorov turbulence. The second
is that the flow is no longer isotropic and hence the effective viscosity should
be a tensor, rather than a scalar.

As a first attempt to explore this possibility
\citet{Penev_Sasselov_Robinson_Demarque_07, Penev_Sasselov_Robinson_Demarque_08}
adapted the \citet{Goodman_Oh_97} perturbative calculation to the
\citet{Robinson_et_al_03} numerical model of stellar convection and found an
asymmetric effective viscosity that scaled linearly with the period of
the external shear. However, their perturbative calculation is applicable only
as long as the forcing period $T$ is small compared to the turnover time of the
largest eddies. In particular the perturbative treatment is not able to
provide the maximum value the effective viscosity reaches and the frequency at
which it reaches it, which is of great importance in calculating tidal
interactions and dissipation of pulsations. 

In this article we use the \citet{Penev_Barranco_Sasselov_08a} spectral anelastic
code to perform a direct calculation of the turbulent dissipation in
a convective zone, by introducing external shear as an extra body force in the
fluid equations. The goal is to investigate the applicability of effective
viscosity as an approximation to the actual turbulent dissipation, and to derive
directly an effective viscosity prescription and compare it against 
the \citet{Goodman_Oh_97} formalism. 

\section{Simulations}
\subsection{Steady State Convection}
The details of the numerical simulation and the equations evolved are presented
in \citet{Penev_Barranco_Sasselov_08a}. We are simulating a rectangular box with
impenetrable, constant temperature top and bottom boundaries (the $\hat{z}$
velocity vanishes and the temperature is held at some constant value at the
top and bottom walls of the box) using the anelastic approximation. 

 The background state, and the parameters with which
all the runs presented in this paper were computed, are the same as the
parameters used for all convectively unstable tests of section 4.1 of
\citet{Penev_Barranco_Sasselov_08a}.
For convenience we remind them here:
\begin{center}
\begin{tabular}{cl}
	$L_x=L_y=L_z=4$ & physical dimensions of the convective box. \\
	$N_x=N_y=N_z=128$ & resolution in each direction\\
	$p_{top}=1.0\times10^5$ & background pressure at the top of the box\\
	$g=2.74$ & gravitational acceleration, in $-\hat{z}$ direction\\
	$C_p=0.21$ & specific heat at constant pressure of the fluid\\
	$R=8.317\times10^{-2}$ & ideal gas constant of the fluid\\
	$T_{low}=10.0$ & temperature at the top boundary of the box\\
	$T_{high}=62.37$ & temperature at the bottom boundary of the box
\end{tabular}
\end{center}

\begin{figure}[tb]
\begin{center}
	\includegraphics[angle=-90, width=0.49\textwidth]{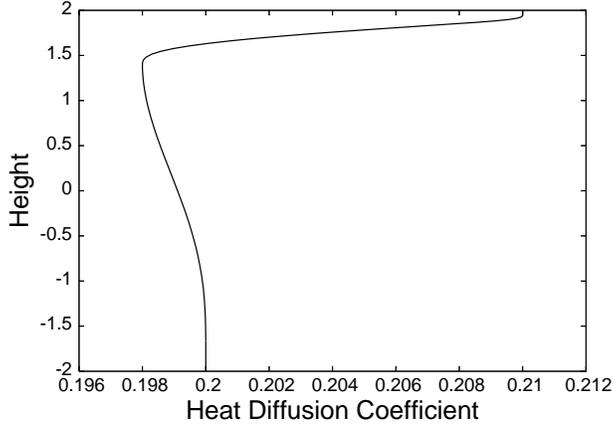}
	\caption{The height dependent heat diffusion coefficient}
	\label{fig: kappa}
\end{center}
\end{figure}

The vertical profile of the heat diffusion coefficient we imposed is presented
in figure \ref{fig: kappa}.

\begin{figure}[tb]
\begin{center}
	\includegraphics[angle=-90, width=0.49\textwidth]{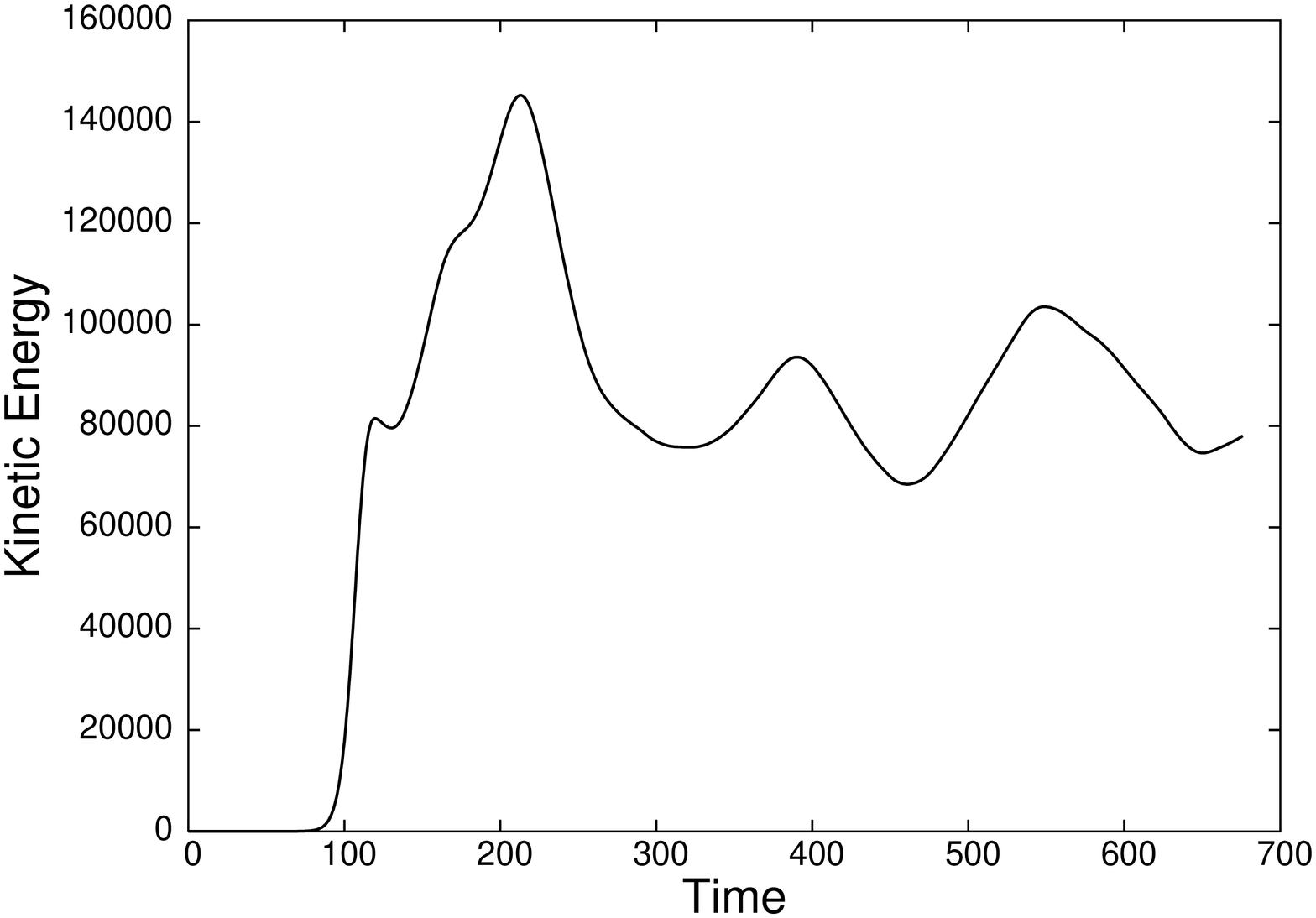}
	\includegraphics[angle=-90, width=0.49\textwidth]{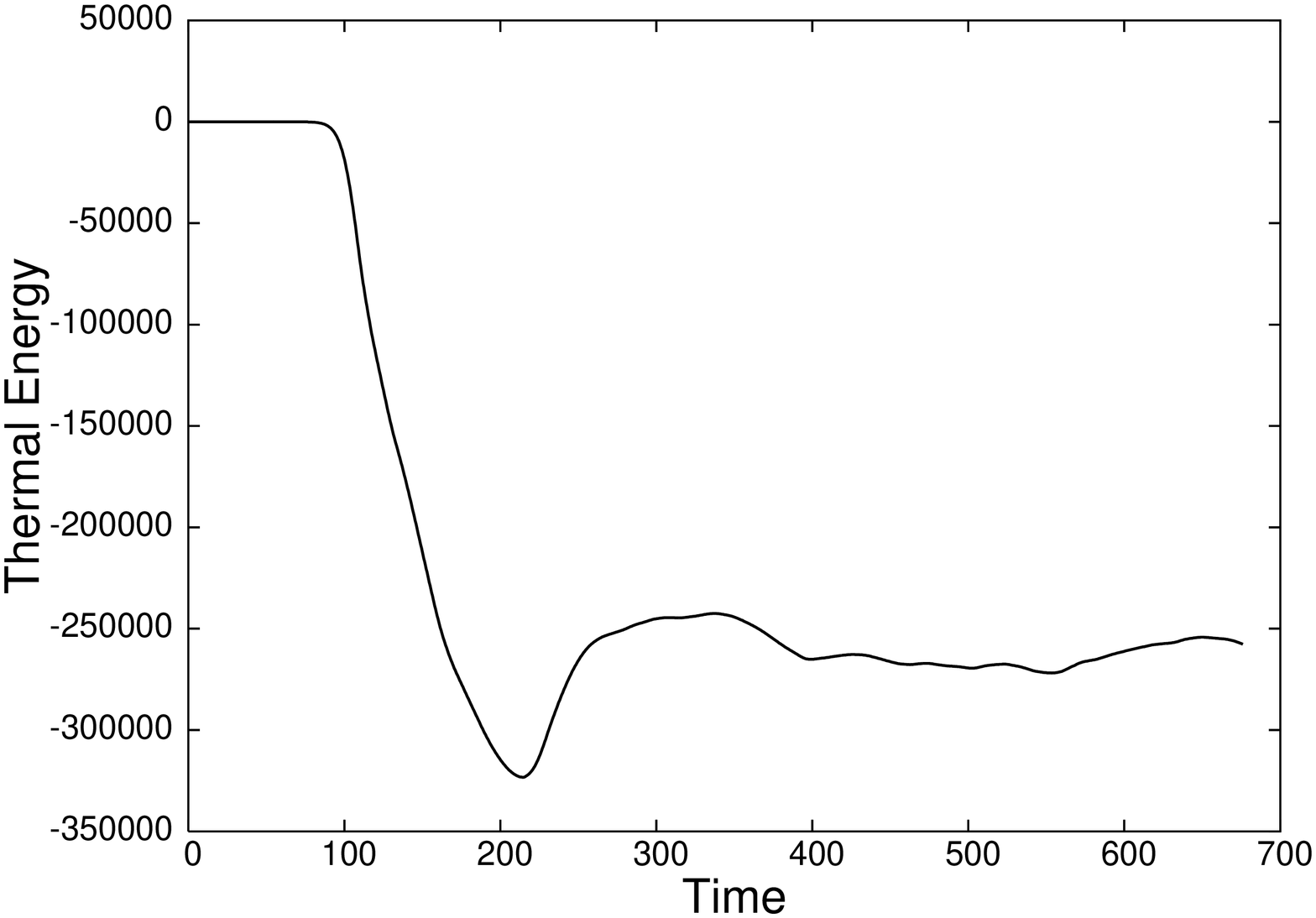}
	\caption{Kinetic (left) and thermal (right) energy content of the
	convective box used as one of the criteria for having reached a steady
	state. We decided steady state was reached for times greater than 400.}
	\label{fig: energy steady}
\end{center}
\end{figure}

\begin{figure}[tbp]
\begin{center}
	\includegraphics[width=0.45\textwidth]{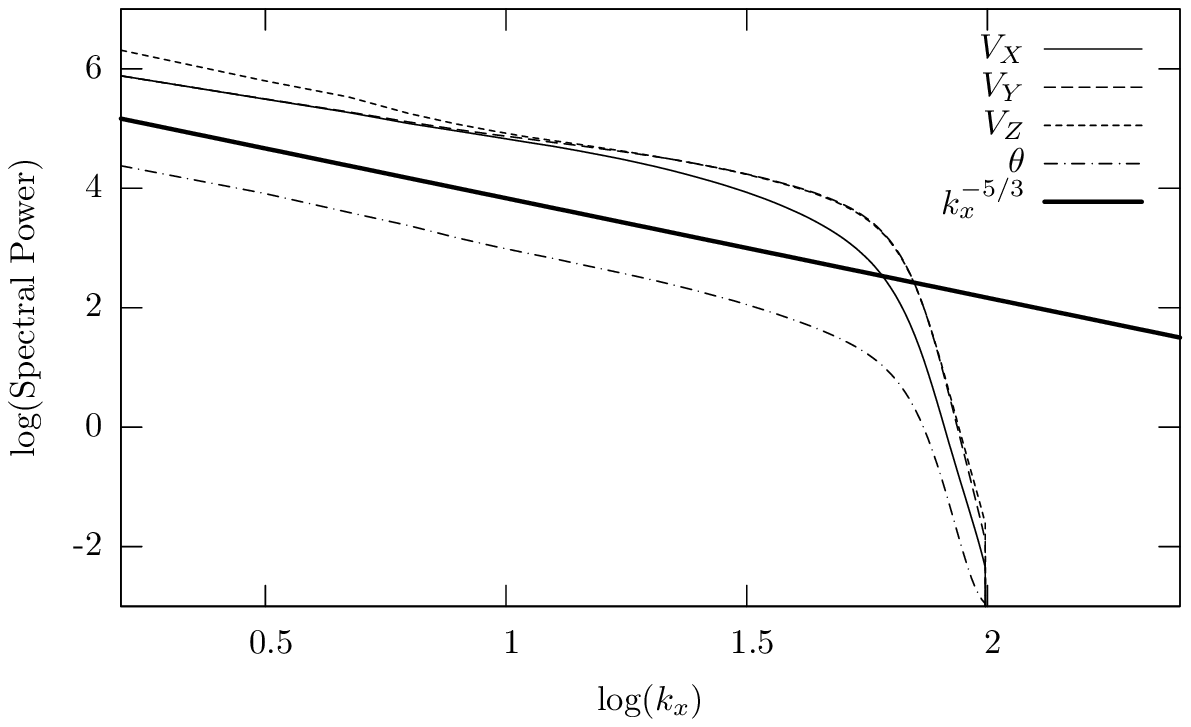}
	\includegraphics[width=0.45\textwidth]{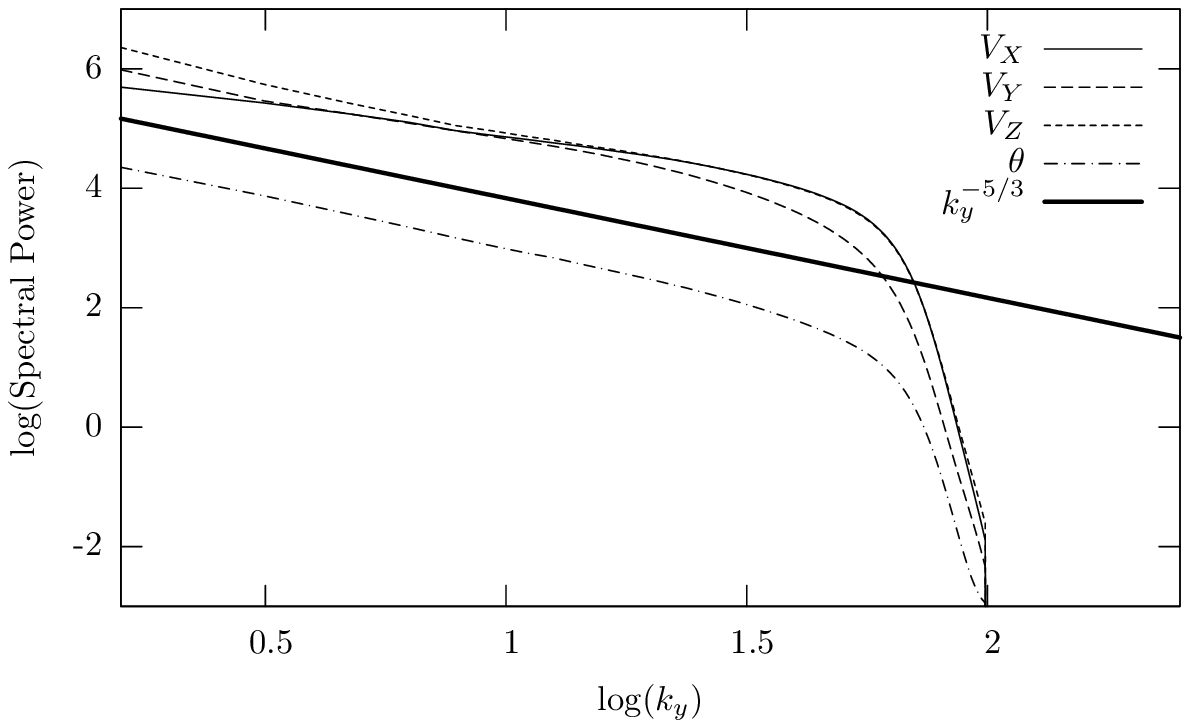}
	\includegraphics[width=0.45\textwidth]{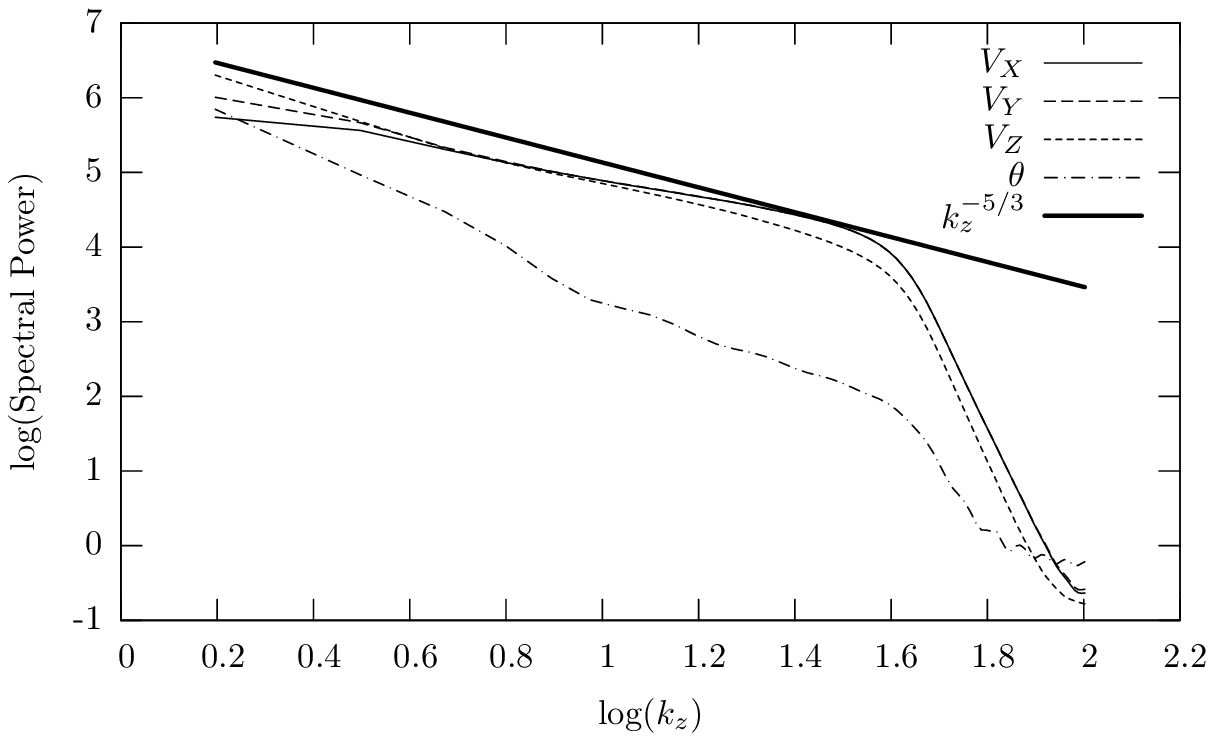}
	\includegraphics[width=0.45\textwidth]{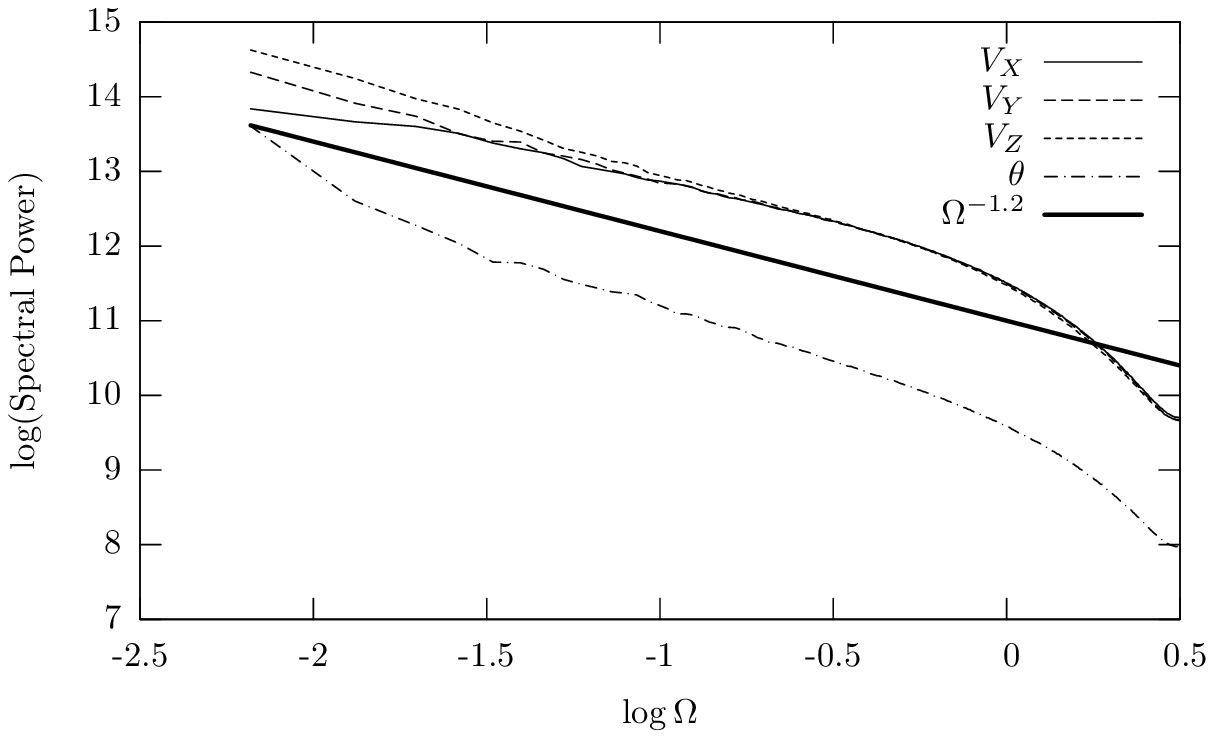}
	\caption{The (x: top left, y: top right, z: bottom left, time: bottom
	right) spectra of the 3 velocity components and the potential 
	temperature. The thick line in the spatial spectra plots corresponds to
	Kolmogorov scaling ($E_k\propto k^{-5/3}$). The thick line in the time
	spectra plot corresponds to the scaling
	\citet{Penev_Barranco_Sasselov_08a} found for the effective viscosity.}
	\label{fig: fff spectra}
\end{center}
\end{figure}

We initialize the box with random entropy fluctuations and let it evolve with
time until a steady state is reached. The criteria we used for concluding a
steady state has been reached were, that the kinetic and thermal energies should
stop drifting systematically, and only exhibit oscillations at the approximate
convective turnover time (see fig. \ref{fig: energy steady}), and that the
spatial spectra of the velocity and potential temperature remain constant to
within a few percent. The steady state Fourier spatial and time spectra are
presented in fig. \ref{fig: fff spectra}.

\subsection{External Forcing}
\label{sec: external_forcing}
After steady state has been reached we introduce external forcing ($\mathbf{f}$) in the
form of a position and time dependent gravitational acceleration in addition to
the already present vertical gravity. So, in short, the anelastic
momentum equation that we evolve is:
\begin{equation}
	\frac{\partial \mathbf{v}}{\partial t} =  
	\mathbf{v}\times\mathbf{\omega}
	- \nabla \widetilde{h} +
	\frac{\widetilde{\theta}}{\bar{\theta}}g\hat{\mathbf{z}}+\mathbf{f},
\end{equation}
where quantities with tilde represent anelastic perturbations to the background
variable (denoted by an over bar), $\mathbf{v}$ is the velocity vector,
$\widetilde{h}\equiv\widetilde{p}/\bar{\rho}+\mathbf{v}^2/2$ is the enthalpy,
$\widetilde{\theta}$ is the perturbation to the background potential
temperature, $\bar{\theta}=\bar{T}\left(p_0/\bar{p}\right)^{R/C_p}$, and
$\mathbf{f}$ is the external forcing. We have investigated the effects of two
forms of forcing:
\begin{enumerate}
	\item $Z$ (height) dependent horizontal forcing with a Gaussian profile:
	\begin{equation}
		\mathbf{f}(z,t)=f_0 e^{-2z^2} \cos\left(\frac{2\pi t}{T}\right)
		\mathbf{\hat{x}}
		\label{eq: f(z)}.
	\end{equation}
	\item $Y$ dependent forcing in the $\hat{x}$ direction, again with a
	Gaussian profile: 
	\begin{equation}
		\mathbf{f}(y)=f_0 e^{-2y^2} \cos\left(\frac{2\pi t}{T}\right)
		\mathbf{\hat{x}}
		\label{eq: f(y)}.
	\end{equation}
\end{enumerate}	

Clearly the amplitude of the forced velocity will be approximately 
$f_0 T/2\pi$, so in order to investigate the period dependence of the
dissipation, we simulated a number of flows with different periods and the
same $f_0 T$. That way, the shear created by the external forcing was the same
for all the flows in the set. We computed two such sets for each forcing case:
one  with $f_0 T=1$ and another with $f_0 T=0.15$. In addition we performed
another set of simulations with fixed period and varying values of $f_0 T$ in
order to investigate the effects of the forcing amplitude on the dissipation
efficiency. Appendix \ref{app: runs} summarizes the runs and the respective
number of time steps simulated for each case.

Notice that the weak forcing cases require a lot more time steps in order to
average out the turbulent noise and allow us to detect the systematic energy
dissipation. Similarly, the $y$ dependent forcing requires longer runs as well,
because in this case, for the same forcing there is less dissipation due to the
anisotropic effective viscosity. 

\begin{figure}[tb]
\begin{center}
	\includegraphics[width=0.45\textwidth]{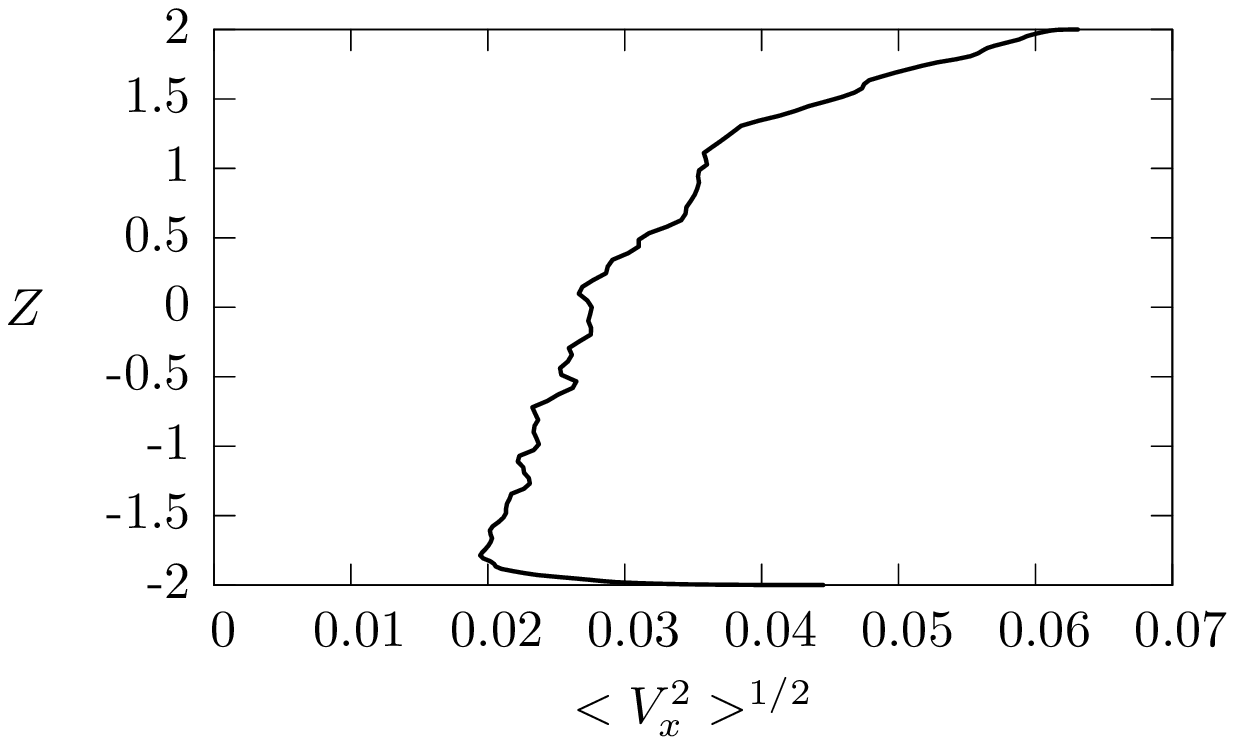}
	\includegraphics[width=0.45\textwidth]{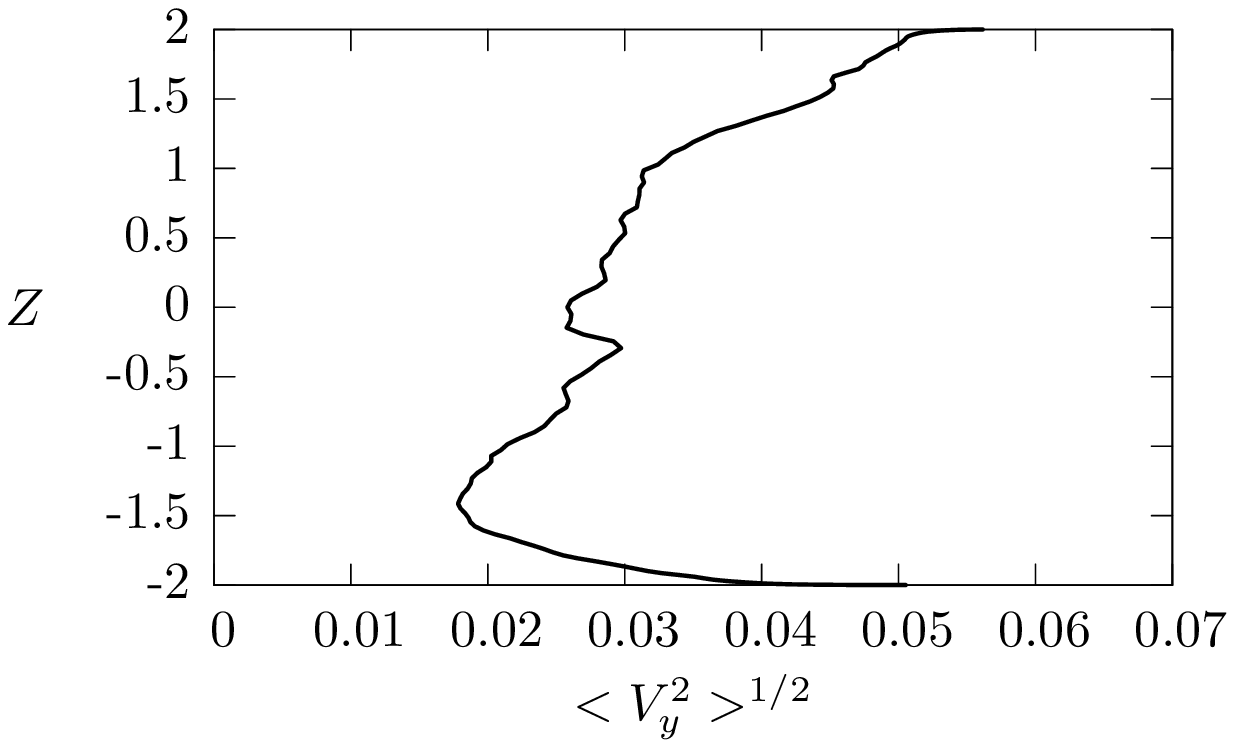}
	\includegraphics[width=0.45\textwidth]{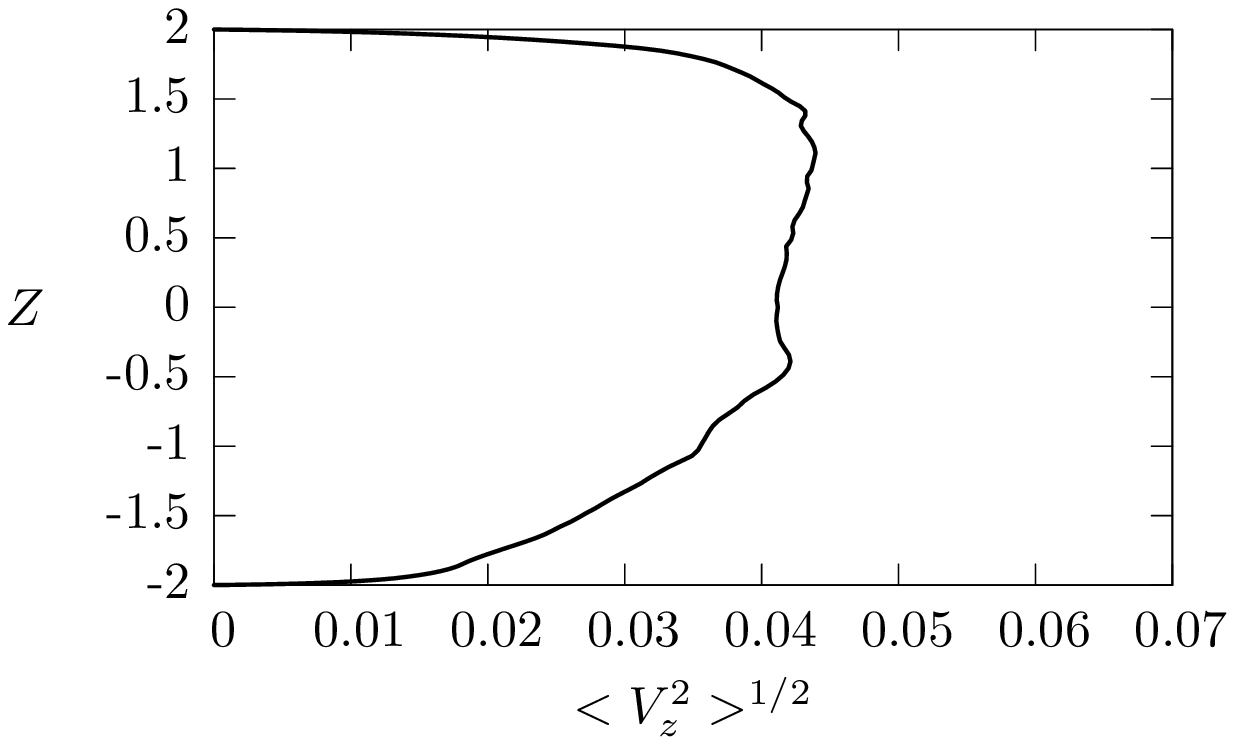}
	\caption{The typical steady state r.m.s. velocities ($v_x$ top left, $v_y$ top
	right, $v_z$ bottom) in the absence of forcing.}
	\label{fig: rms_v}
\end{center}
\end{figure}

For the strong forcing case, we expect the maximal central velocities to reach
$\max v_x=1/2\pi\approx0.16$, and for the weak forcing $\max
v_x=0.15/2\pi\approx0.024$. For comparison, in our convective box the
speed of sound varies between $1.2$ at the top of the box and $2.9$ at the
bottom and the typical r.m.s. velocities are between $0.02$ and $0.04$ (see
figure \ref{fig: rms_v}), except very near the top and bottom of the box where
the collision of the vertical flow with the impenetrable boundaries results in
higher horizontal velocities.

In the case of tides in binary stellar systems, the velocities excited
by the external forcing are small compared to the typical convective velocities,
however, performing numerical simulations with forcing small enough to ensure
that this holds will be prohibitive in terms of computational time,
because it will require simulations for excessive number of time steps to make
sure the dissipation is noticeable among the fluctuations due to the turbulence
(see section \ref{sec: direct dissipation}). 

\section{Results}
\label{sec: dv results}
\subsection{Anisotropic Viscosity}
The viscosity in this work, as in \citet{Penev_Barranco_Sasselov_08a} is
described by a rank--4 tensor ($K_{ijmn}$), which
gives the relation between the strain rate $e_{ij}\equiv1/2(\partial_i v_j +
\partial_j v_i)$ and the viscous stress $\sigma_{ij}$:
\begin{equation}
	\sigma_{ij}=K_{ijmn} e_{mn},
\end{equation}
where summation over repeated indices is assumed.

This tensor must obey a set of symmetries in order to avoid infinite torques. In
addition we expect it to be symmetric with respect of rotations and reflections
around the $\hat{z}$ axis. With all these symmetries $K_{ijmn}$ can be shown to
have only six independent components. Of those, the particular forms of forcing
described in section \ref{sec: external_forcing} probe only two families:
\begin{equation}
	\begin{array}{r@{=}l}
	K_{1221}=K_{2121}=K_{2112}&	K_{1212}, \\
	\left.\begin{array}{r}
	K_{3131}=K_{3113}=K_{1331}=K_{2323}\\
	K_{3232}=K_{3223}=K_{2332}
	\end{array}\right\}	&	K_{1313},\\
	\end{array}
\end{equation}
which will be referred to as $K_{1212}$ and $K_{1313}$ from now on.

\subsection{Direct Viscosity Calculation}
\label{sec: direct dissipation}
To simplify the discussion we define upfront the following quantities for the
$z$ dependent forcing:
\begin{eqnarray}
	C_{xz}(z)&=&\sum_{i,j,w} v_x(x_i, y_j, z, t_w) \cos(\frac{2\pi t_w}{T})
	\label{eq: Cxz(z)},\\
	\mathrm{and}\quad S_{xz}(z)&=&\sum_{i,j,w} v_x(x_i, y_j, z, t_w)
	\sin(\frac{2\pi t_w}{T}),
	\label{eq: Sxz(z)}
\end{eqnarray}
and for the $y$ dependent forcing:
\begin{eqnarray}
	C_{xy}(y,z)&=&\sum_{i,w} v_x(x_i, y, z, t_w) \cos(\frac{2\pi t_w}{T})
	\label{eq: Cxy(z)},\\
	\mathrm{and}\quad S_{xy}(y,z)&=&\sum_{i,w} v_x(x_i, y, z, t_w)
	\sin(\frac{2\pi t_w}{T}),
	\label{eq: Sxy(z)}
\end{eqnarray}
where $x_i$ and $y_j$ are the locations of the $x$ and $y$ collocation grid
points, and $t_w$ are the times at which we have sampled the velocity field. We
evaluate the sum over an integer number of forcing periods $T$.

\subsubsection{Depth Dependence of the Effective Viscosity}
We use pre-determined functions of depth, up to a normalization constant,
for the effective viscosity coefficients that we need, namely:
\begin{eqnarray}
	K_{1313}&=&K^0_{1313} \left<v_z^2\right>^{1/2} H_p
	\label{eq: nu_xz_form},\\
	\mathrm{and}\quad K_{1212}&=&K^0_{1212}
	\left<\frac{v_x^2+v_y^2}{2}\right>^{1/2} H_p.
	\label{eq: nu_xy_form}
\end{eqnarray}
This is the same scaling that we used in presenting the perturbative result
\citet{Penev_Barranco_Marcus_08a},
with the only difference that only the velocity in the direction of the external
shear is used as the velocity scale. In equation \ref{eq: nu_xy_form}
we average together both horizontal components of the velocity, because on
average there should be no physical difference between the two. In practice, the
different velocity scaling makes little difference, since as we can see from fig.
\ref{fig: rms_v}, away from the boundaries all components of the velocity behave
alike, except for the fact that $v_z$ tends to be larger. So, using the full
r.m.s. velocity instead of only one component, just leads to smaller values of
the normalization constants $K^0_{1313}$ and $K_{1212}^0$.

\subsubsection{Fitting the Spatial Dependence of the Dissipation}
\label{sec: fitting}
We would like to verify the applicability of the effective viscosity framework
to the problem of turbulent dissipation, by showing that substituting
the turbulent flow with a simple viscosity is able to capture not only the total
amount of energy dissipated, but also the momentum transport, or in other words
the spatial dependence of this dissipation. 

For the $z$ dependent $x$ forcing, we would like to show that the work
per unit mass done by the forcing on the flow at each depth:
\begin{equation}
	W_{xz}^{turb}(z)\equiv f_0 e^{-2z^2} C_{xz}(z),
	\label{eq: Wxz_turb}
\end{equation}
matches the energy that would be transported and dissipated out of that depth by
an assumed effective viscosity:
\begin{equation}
	\begin{array}{rcl}
	\displaystyle{
	W_{xz}^{visc}(z)}&\equiv&\displaystyle{
		\frac{1}{2}K_{1313}(z)\left[C_{xz}(z) C_{xz}''(z) + 
		S_{xz}(z) S_{xz}''(z)\right] + {}}\\
	&&\displaystyle{{}+\frac{1}{2}\left(K_{1313}'(z)+\frac{d\ln
	\bar{\rho}}{dz}K_{1313}(z)\right)\left[
	C_{xz}(z) C_{xz}'(z) + S_{xz}(z) S_{xz}'(z)\right],}
	\end{array}
	\label{eq: Wxz_visc}
\end{equation}
where primes denote derivatives with respect to $z$ and $\bar{\rho}$ is the
background density profile. 

For the $y$ dependent $x$ forcing, we would like to show that the work
per unit mass done by the forcing on the flow at each $y$ plane:
\begin{equation}
	W_{xy}^{turb}(y)\equiv \frac{f_0}{N} e^{-2y^2}
	\int_{z_{min}}^{z_{max}} \bar{\rho}(z) C_{xy}(y,z) dz,
	\label{eq: Wxy_turb}
\end{equation}
matches the energy that would be transported and dissipated out of that plane by
an assumed effective viscosity:
\begin{equation}
	W_{xy}^{visc}(y)\equiv\frac{1}{2N}
		\int_{z_{min}}^{z_{max}} \bar{\rho}(z)
		K_{1212}(z)\left[
			C_{xy}(y,z) C_{xy}''(y,z) + 
			S_{xy}(y,z) S_{xy}''(y,z)
		\right]dz,
	\label{eq: Wxy_visc}
\end{equation}
where now primes denote derivatives with respect to $y$, $z_{min}$ and
$z_{max}$ are the minimal and maximal depth respectively that we want to include
in the fit and $N\equiv\int_{z_{min}}^{z_{max}} \bar{\rho}(z)dz$. The reason we do not want to include the entire simulated domain is
that near the boundaries the flow is strongly affected by the impenetrable top
and bottom walls and is thus non-physical.

\begin{figure}[tbp]
\begin{center}
	\includegraphics[width=0.45\textwidth]{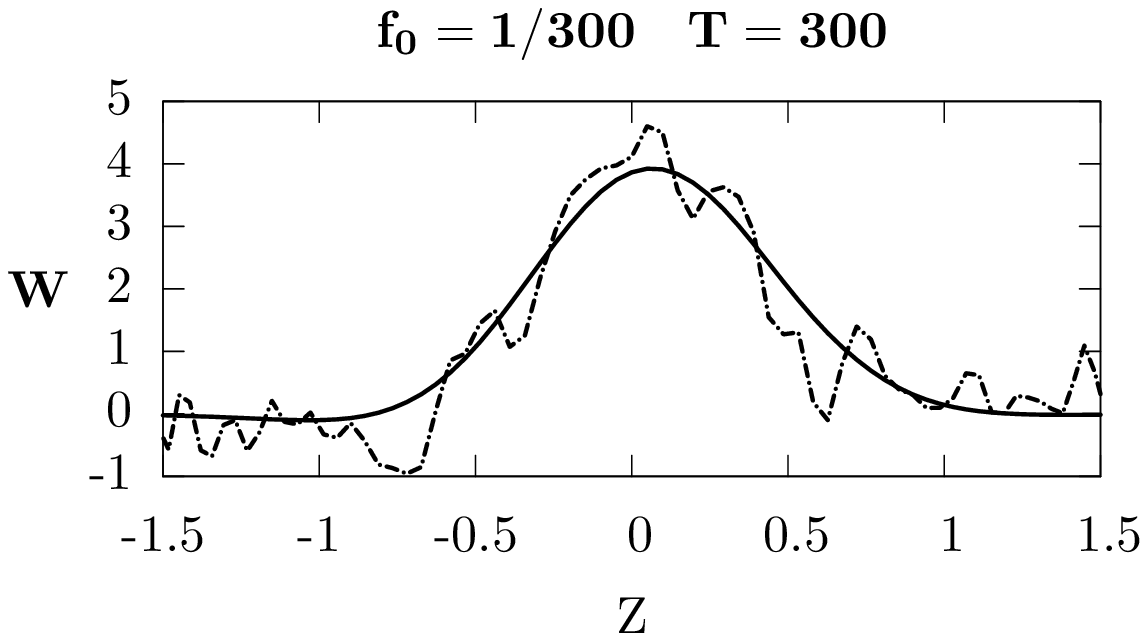}
	\includegraphics[width=0.45\textwidth]{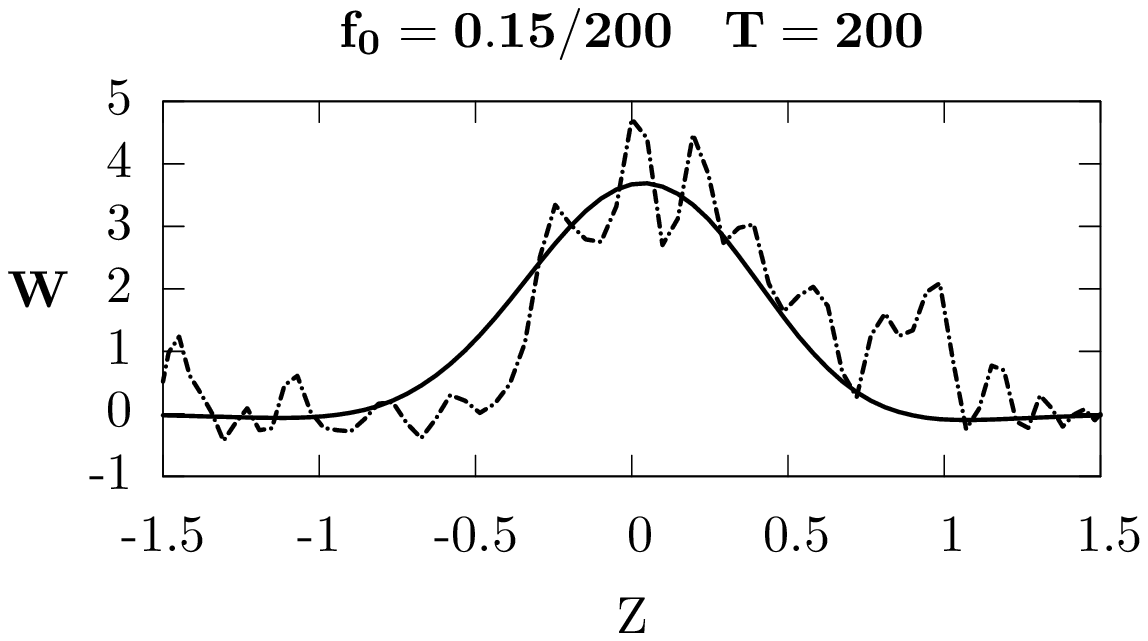}
	\includegraphics[width=0.45\textwidth]{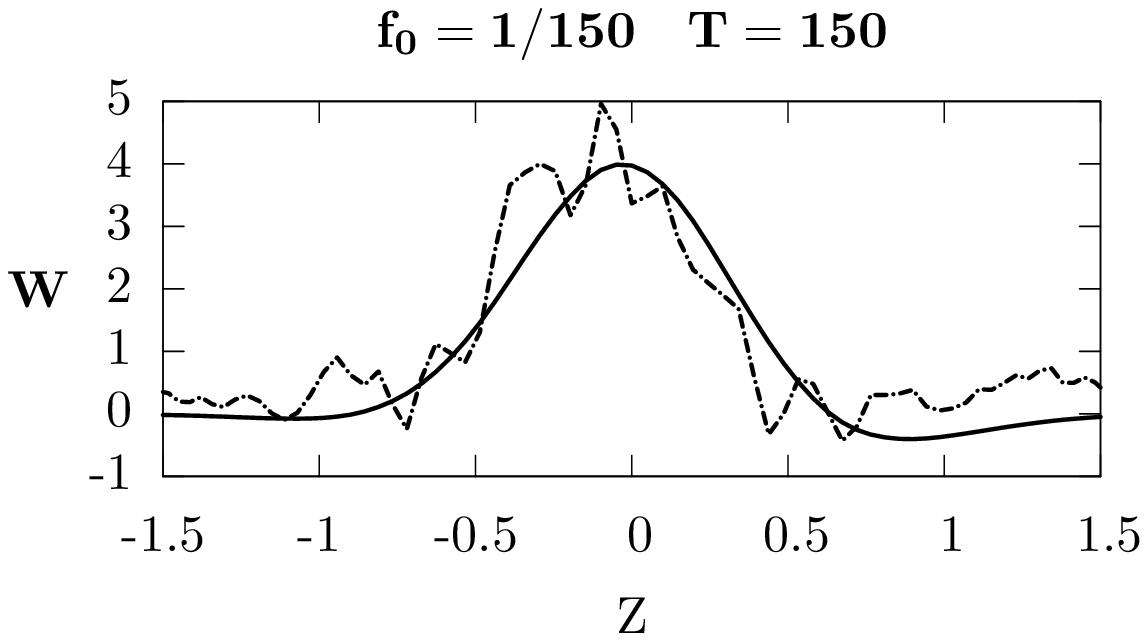}
	\includegraphics[width=0.45\textwidth]{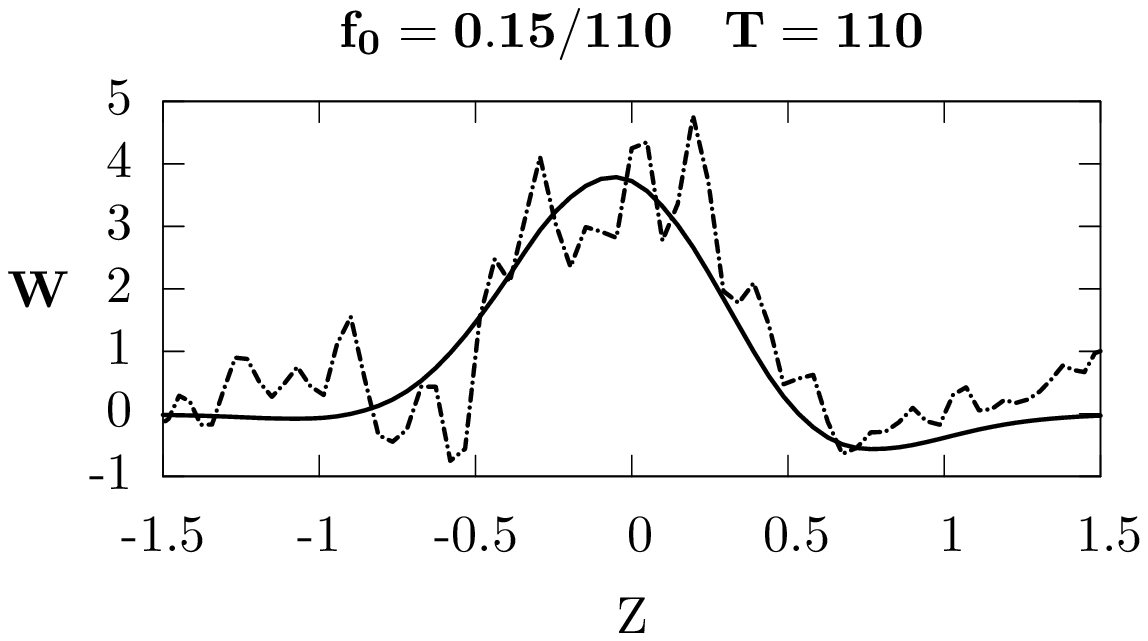}
	\includegraphics[width=0.45\textwidth]{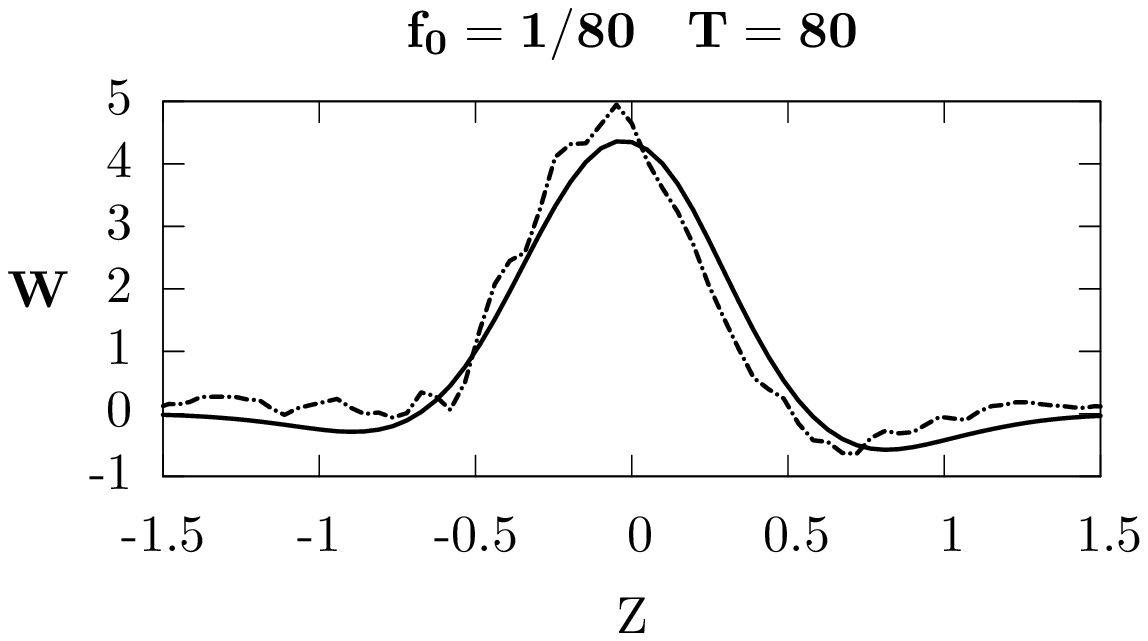}
	\includegraphics[width=0.45\textwidth]{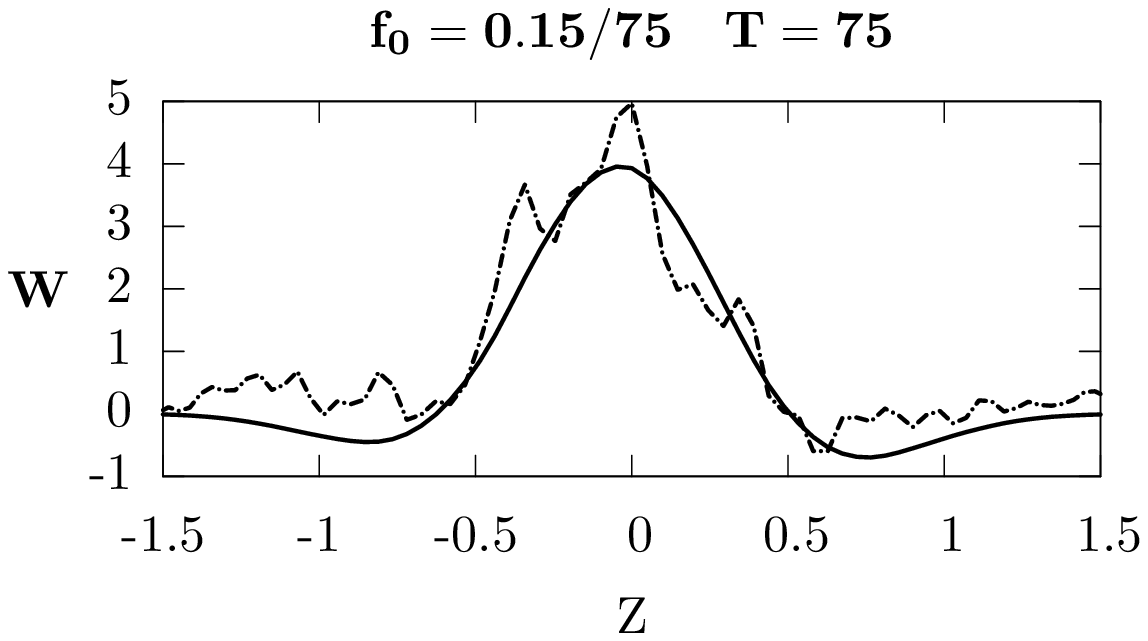}
	\includegraphics[width=0.45\textwidth]{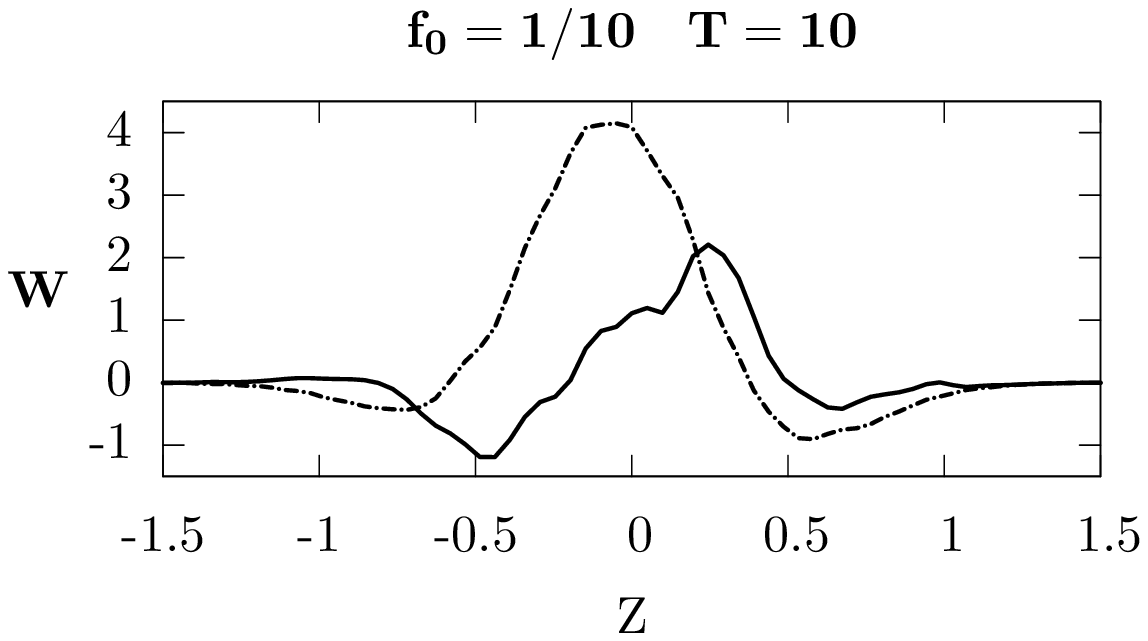}
	\includegraphics[width=0.45\textwidth]{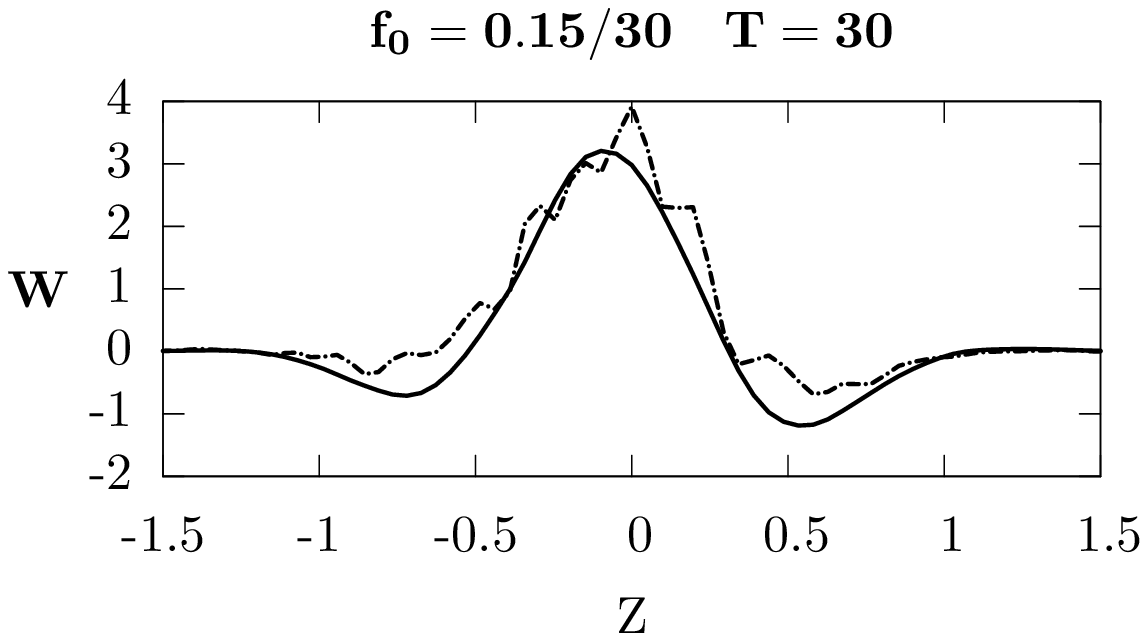}
	\caption{Typical least squares fits between $W_{xz}^{turb}$ (solid
	curves) and $W_{xz}^{visc}$ (dashed curves) for the two sets of
	simulations with $z$ dependent $x$ forcing with fixed amplitude and
	variable period. Left: strong forcing, right: weak forcing. In each plot
	$W_{xz}^{turb}$ and $W_{xz}^{visc}$ have been scaled in order to make
	their maximum values be of order few.}
	\label{fig: XZ_W_fit}
\end{center}
\end{figure}

\begin{figure}[tbp]
\begin{center}
	\includegraphics[width=0.45\textwidth]{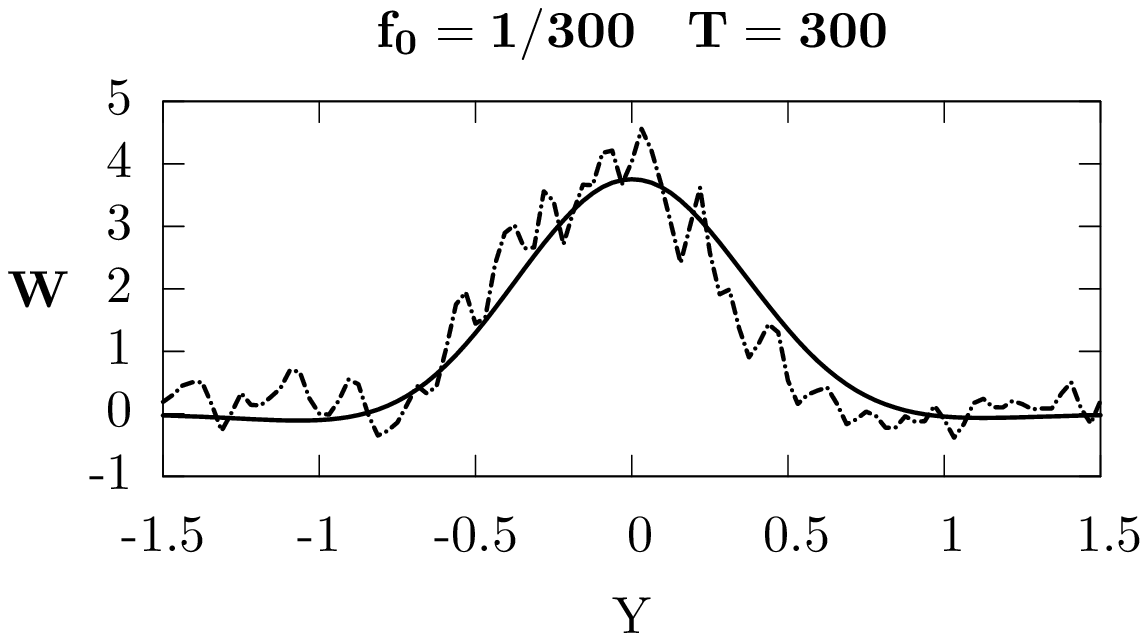}
	\includegraphics[width=0.45\textwidth]{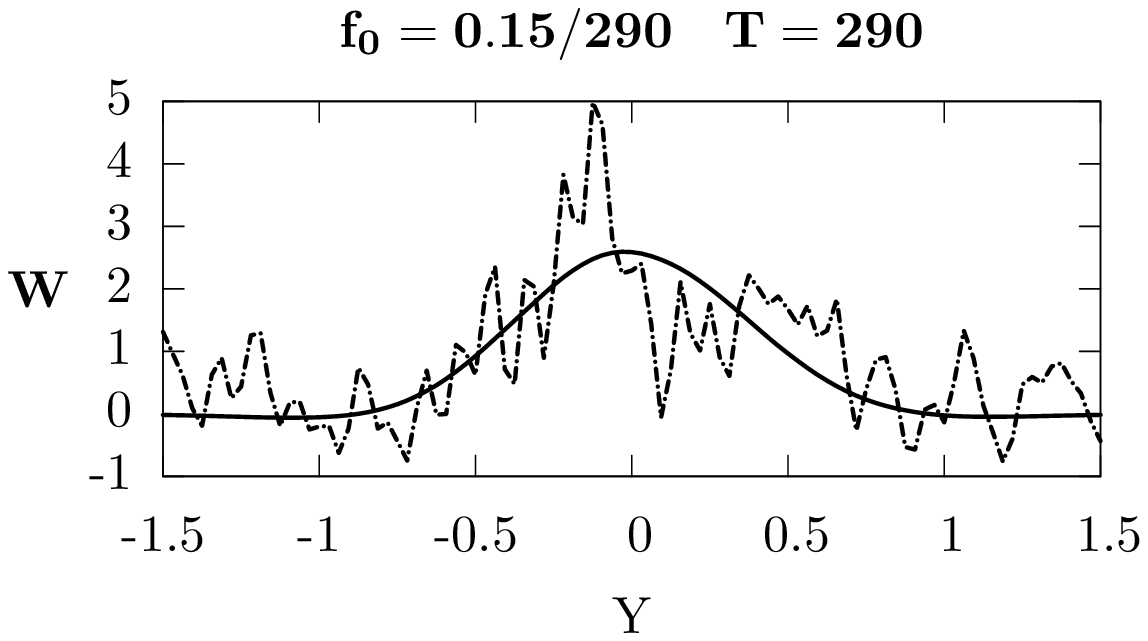}
	\includegraphics[width=0.45\textwidth]{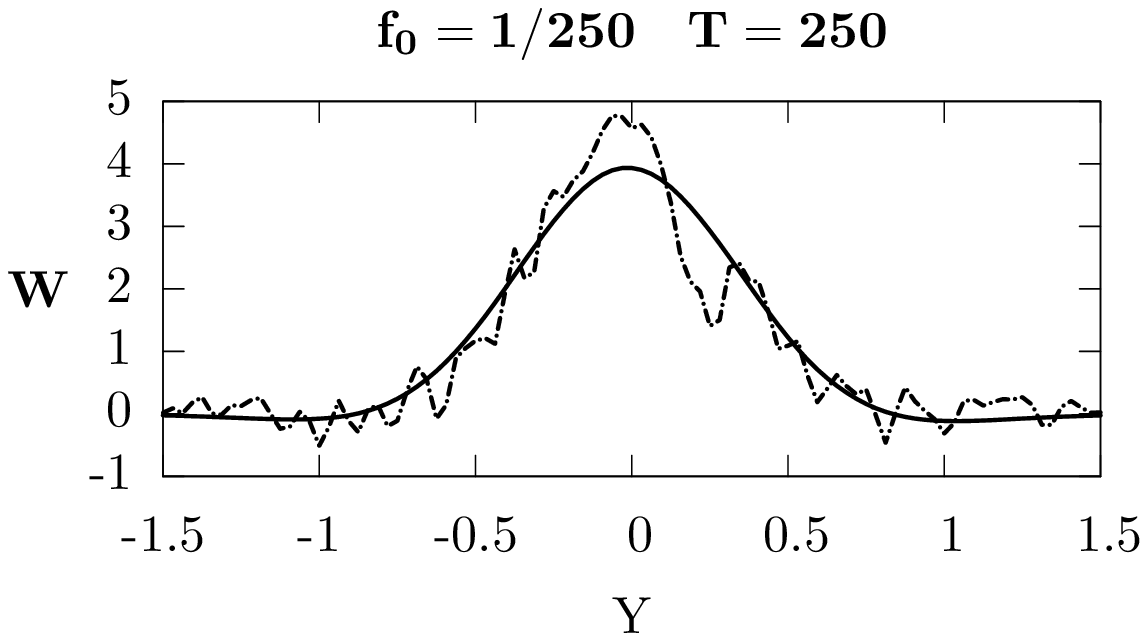}
	\includegraphics[width=0.45\textwidth]{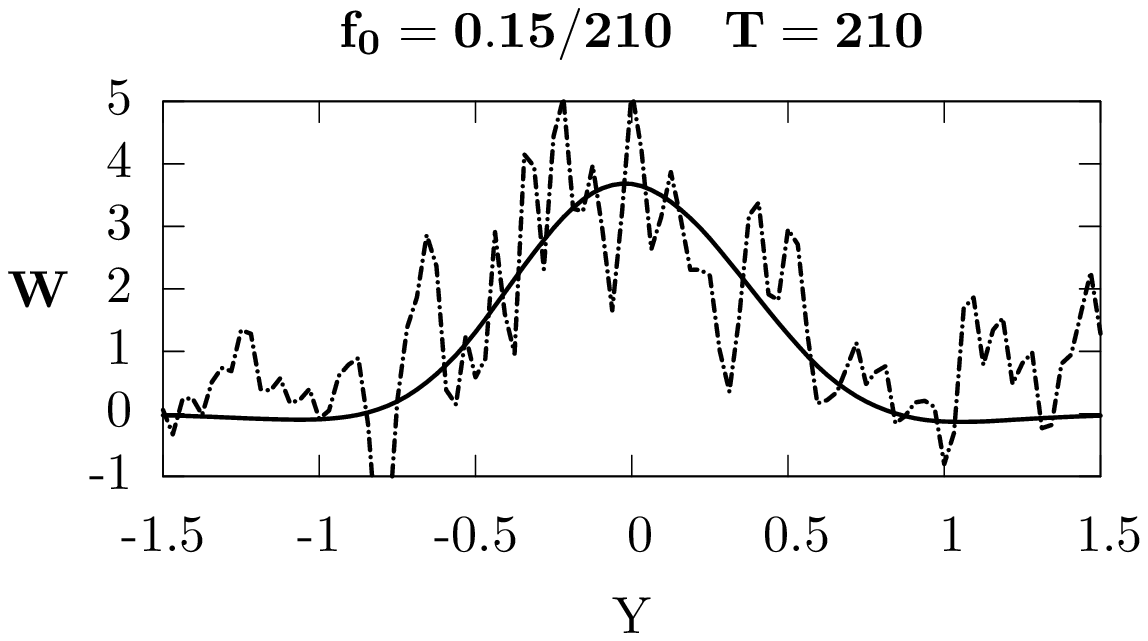}
	\includegraphics[width=0.45\textwidth]{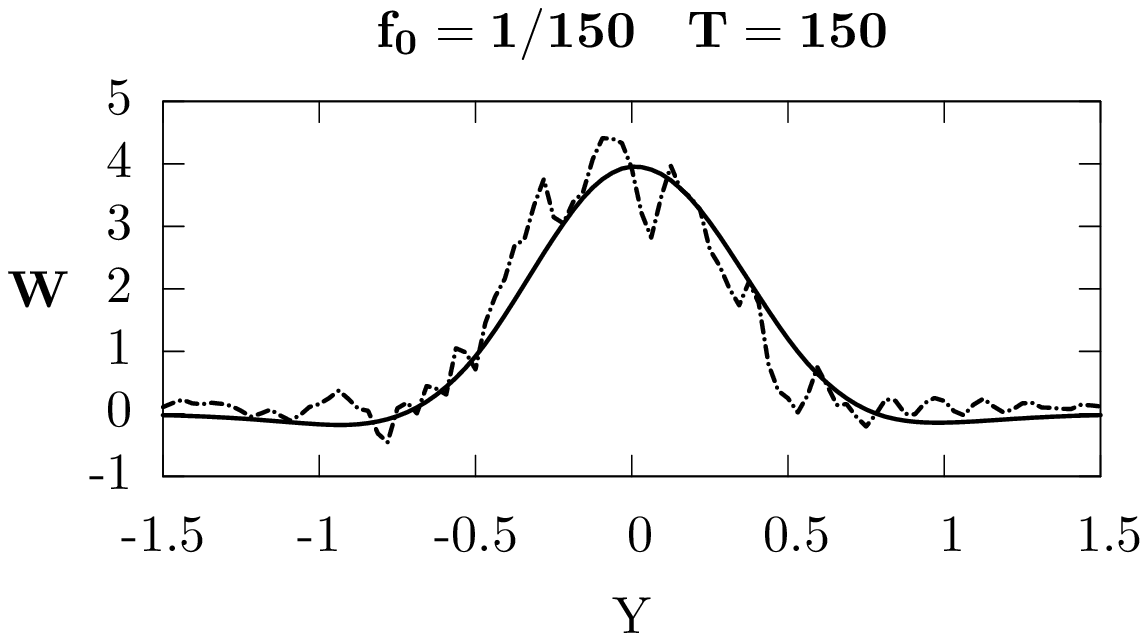}
	\includegraphics[width=0.45\textwidth]{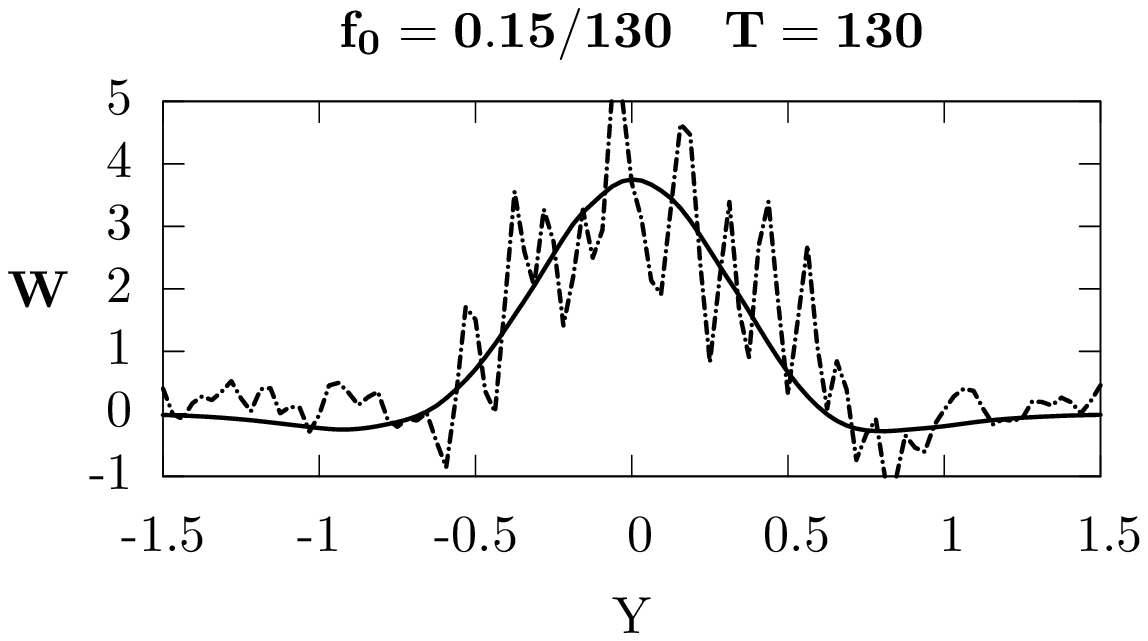}
	\includegraphics[width=0.45\textwidth]{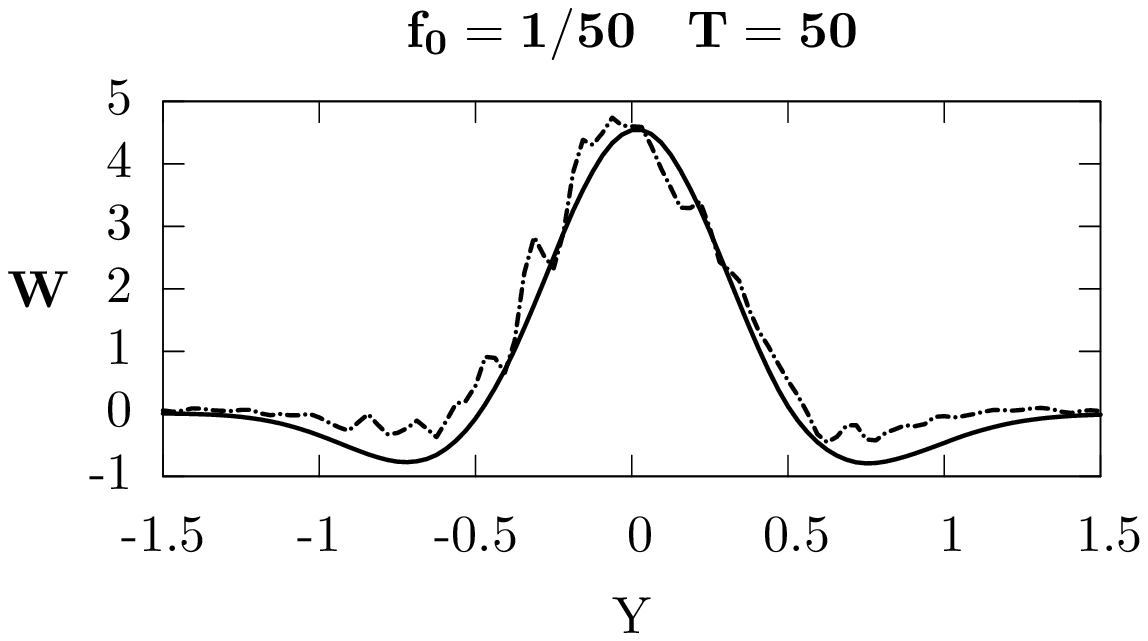}
	\includegraphics[width=0.45\textwidth]{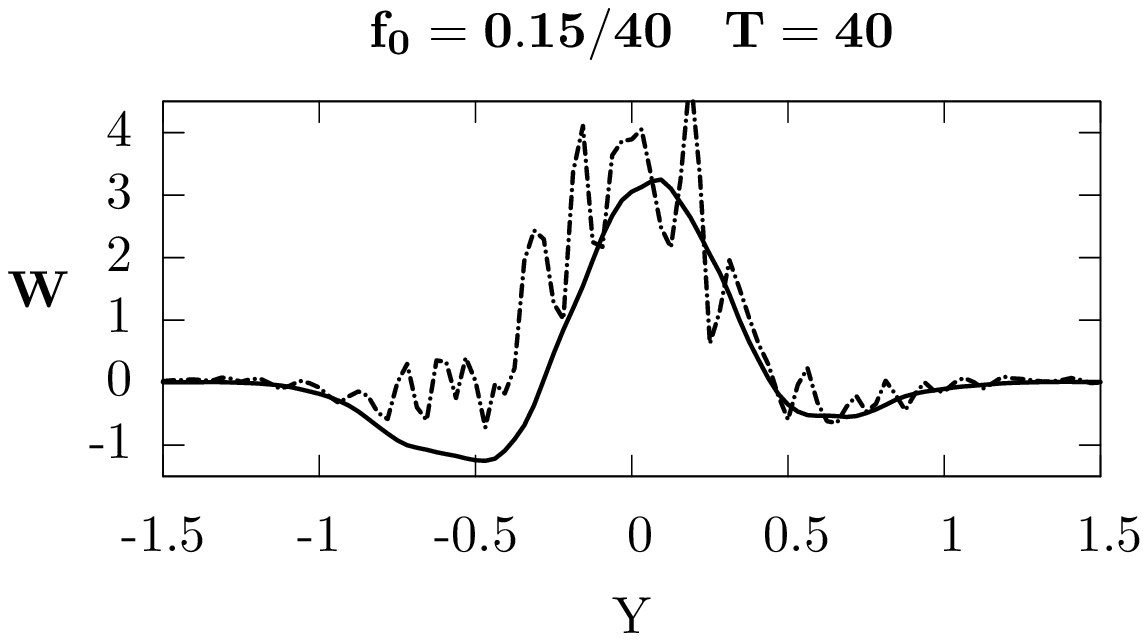}
	\caption{Typical least squares fits between $W_{xy}^{turb}$ (solid
	curves) and $W_{xy}^{visc}$ (dashed curves) for the two sets of
	simulations with $y$ dependent $x$ forcing with fixed amplitude and
	variable period. Left: strong forcing, right: weak forcing. In each plot
	$W_{xy}^{turb}$ and $W_{xy}^{visc}$ have been scaled in order to make
	their maximum values be of order few.}
	\label{fig: XY_W_fit}
\end{center}
\end{figure}

\begin{figure}[tbp]
\begin{center}
	\includegraphics[width=0.45\textwidth]{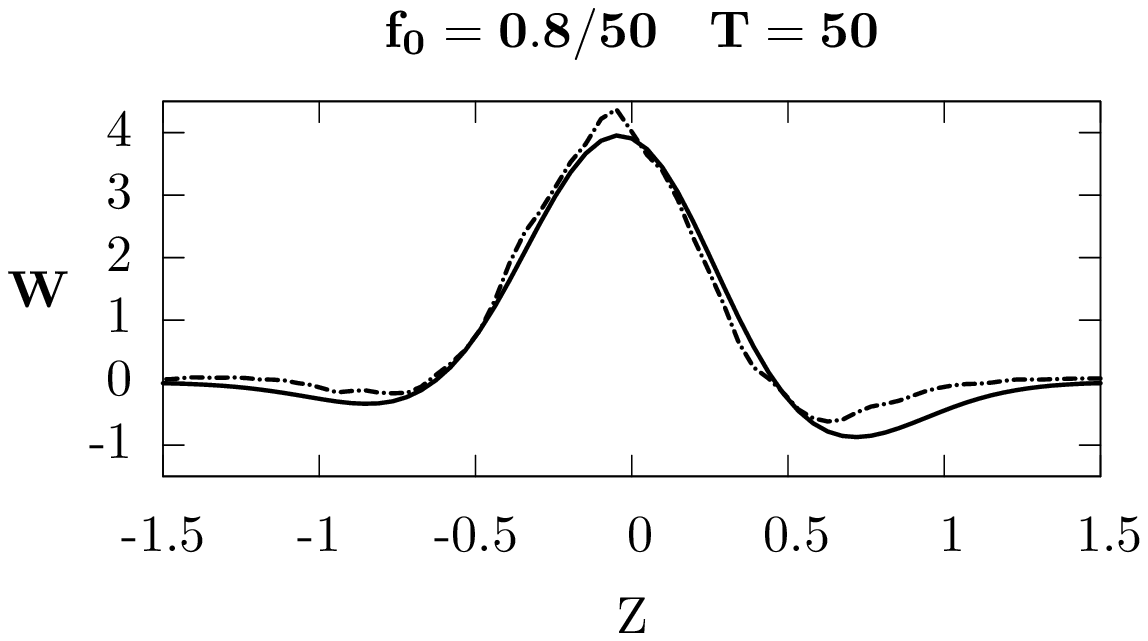}
	\includegraphics[width=0.45\textwidth]{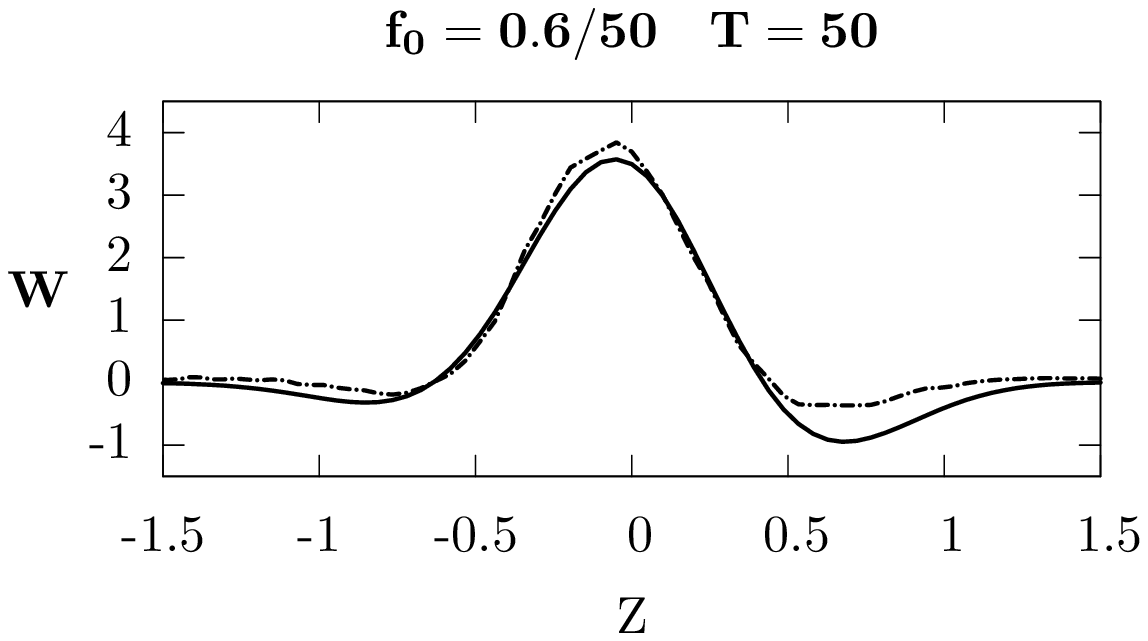}
	\includegraphics[width=0.45\textwidth]{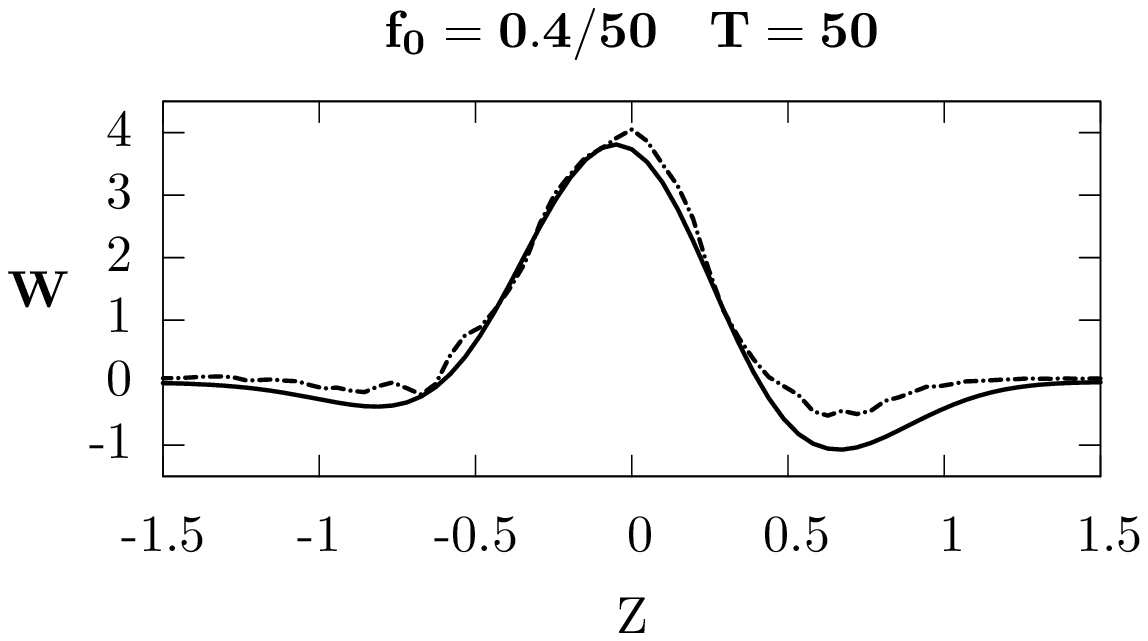}
	\includegraphics[width=0.45\textwidth]{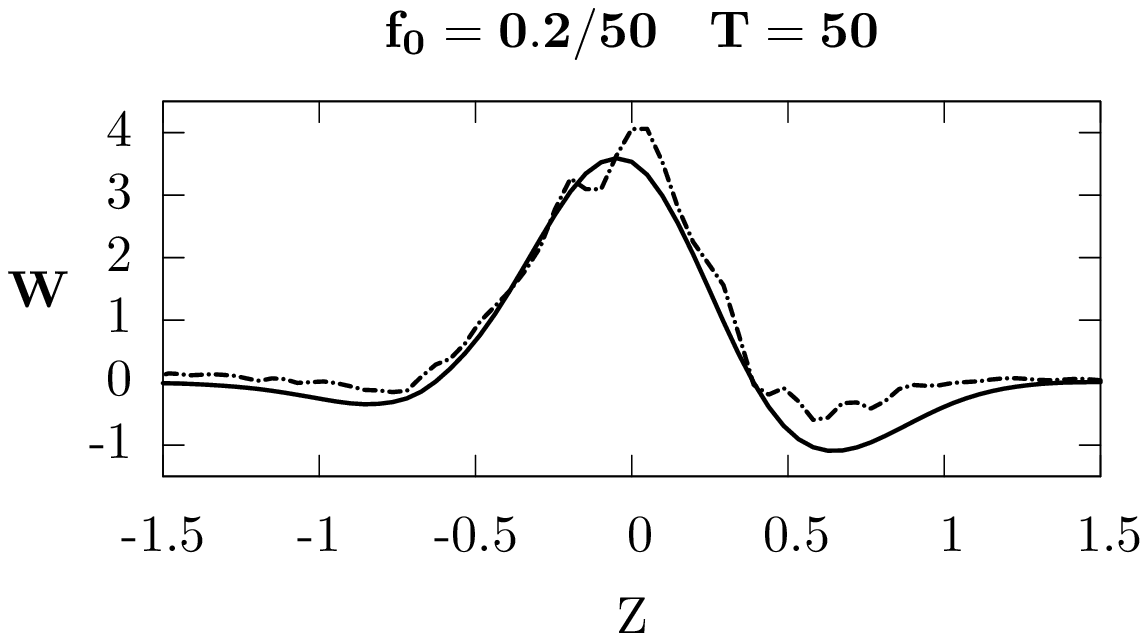}
	\caption{The least squares fits between $W_{xz}^{turb}$ (solid
	curves) and $W_{xz}^{visc}$ (dashed curves) for the set of
	simulations with $z$ dependent $x$ forcing with fixed period and
	variable amplitude. In each plot
	$W_{xz}^{turb}$ and $W_{xz}^{visc}$ have been scaled in order to make
	their maximum values be of order few.} 
	\label{fig: Amp_W_fit}
\end{center}
\end{figure}

We find the values of $K^0_{1313}$ and $K^0_{1212}$ by least squares fitting of
$W_{xz}^{visc}$ to $W_{xz}^{turb}$ in the range $z_{min}<z<z_{max}$ and 
$W_{xy}^{visc}$ to $W_{xy}^{turb}$ in the range $-L_y/2<y<L_y/2$
respectively. Clearly, the presence of the turbulence will cause random
fluctuations in the velocity profile which should average out if we combine a
large enough number of time steps in evaluating the quantities $C_{xz}$, 
$C_{xy}$, $S_{xz}$ and $S_{xy}$. These fluctuations are highly
amplified when we estimate second derivatives of those quantities, so 
$W_{xz}^{visc}$ and $W_{xy}^{visc}$ suffer much more than $W_{xz}^{turb}$ and
$W_{xy}^{turb}$. Some representative fitted curves for each series of runs are
shown in figures \ref{fig: XZ_W_fit}, \ref{fig: XY_W_fit} and 
\ref{fig: Amp_W_fit}. 

We see that, except for one case, the two curves match closely, which shows that
the effective viscosity assumption captures the effects of turbulent dissipation
to a good degree. The $T=10$ curves for the strong $z$ dependent $x$ forcing 
case do not match well at all. The bad fit is due to the fact that if there is
any dissipation present it is undetectable from within the turbulent noise. 

The
reason for the strongly suppressed dissipation is that for such short periods
the corresponding eddy sizes are too small to be reliably simulated at our current
resolution. This fact points to the Goldreich et. al. picture, that the
dissipation is dominated by eddies with periods close to close that of the
external forcing ($T$), as opposed to the Zahn picture in which the largest
eddies are always the most important, since in the latter case we would expect
the linear scaling to continue to arbitrarily short periods.

\subsubsection{Matching Deposited to Dissipated Power}
\label{sec: calc}
An alternative way to get a value for the scaling constants $K_{1313}^0$ and
$K_{1212}^0$ is to equate the overall power deposited into the box by the
external forcing:
\begin{eqnarray}
	\dot\varepsilon_{xz}^{ext}=\int_{-z_{min}}^{z_{max}} 
		\bar{\rho}(z) W_{xz}^{turb}(z) dz\quad&,
	\mathrm{for}\quad&K_{1313}^0\\
	\label{eq: xz epsilon ext}
	\dot\varepsilon_{xy}^{ext}=N\int_{-L_y/2}^{L_y/2}
		W_{xy}^{turb}(y) dy\quad&,
	\mathrm{for}\quad&K_{1212}^0,
	\label{eq: xy epsilon ext}
\end{eqnarray}
to the value that the effective viscosity dissipates out of the box:
\begin{eqnarray}
	\dot\varepsilon_{xz}^{visc}=\int_{-z_{min}}^{z_{max}} \bar{\rho}(z)
	\widetilde{W}_{xz}^{visc}(z) dz&\quad\mathrm{for}\quad&K_{1313}^0,\\
	\label{eq: xz epsilon visc}
	\dot\varepsilon_{xy}^{visc}=\int_{-L_y/2}^{L_y/2} 
	\widetilde{W}_{xy}^{visc}(y) dy&\quad\mathrm{for}\quad&K_{1212}^0
	\label{eq: xy epsilon visc}.
\end{eqnarray}
Where we have defined: 
\begin{eqnarray}
	\widetilde{W}_{xz}^{visc}(z)&\equiv&K_{1313}(z)\left[
	\left(\frac{dC_{xz}}{dz}\right)^2 + 
	\left(\frac{dS_{xz}}{dz}\right)^2 + 
	\right],\\
	\widetilde{W}_{xy}^{visc}(y)&\equiv&\int_{z_{min}}^{z_{max}}
	\bar{\rho}(z)K_{1212}(z)\left[
	\left(\frac{\partial C_{xy}}{\partial y}\right)^2  +
	\left(\frac{\partial S_{xy}}{\partial y}\right)^2\right]dz.
\end{eqnarray}

Note that in this case we do not expect to match the spatial dependence, only 
the overall rate. The reason for this, is that viscous forces redistribute the
energy in the box as well as dissipate it. So we cannot use these quantities to
examine the applicability of the effective viscosity assumption.

Because evaluating $\widetilde{W}_{xz}^{visc}(z)$ and
$\widetilde{W}_{xy}^{visc}(y)$ requires only first derivatives of the 
sine and cosine velocity components, they suffer significantly less
from the turbulent noise than ${W}_{xz}^{visc}(z)$ and $W_{xy}^{visc}(y)$ 
from section \ref{sec: fitting}. On the other hand, some energy transfer
inevitably occurs near the top and bottom boundaries, which is excluded from
evaluating the $z$ integrals. For the case of $z$ dependent forcing this is a
small amount, since the forcing in that part of the box is very small by
design. However, for the $y$ dependent forcing significant energy transfer
does occur in those regions, which could bias the estimated viscosities toward
lower values (see the last paragraph of section \ref{sec: comparison}).

\begin{figure}
\begin{center}
	\includegraphics[width=0.45\textwidth]{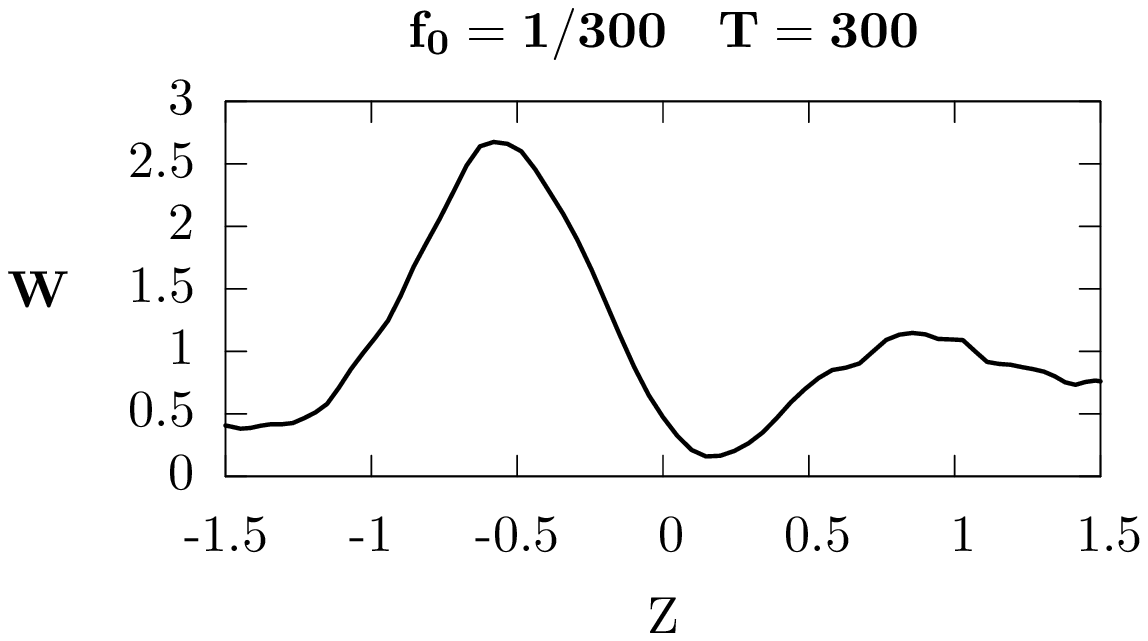}
	\includegraphics[width=0.45\textwidth]{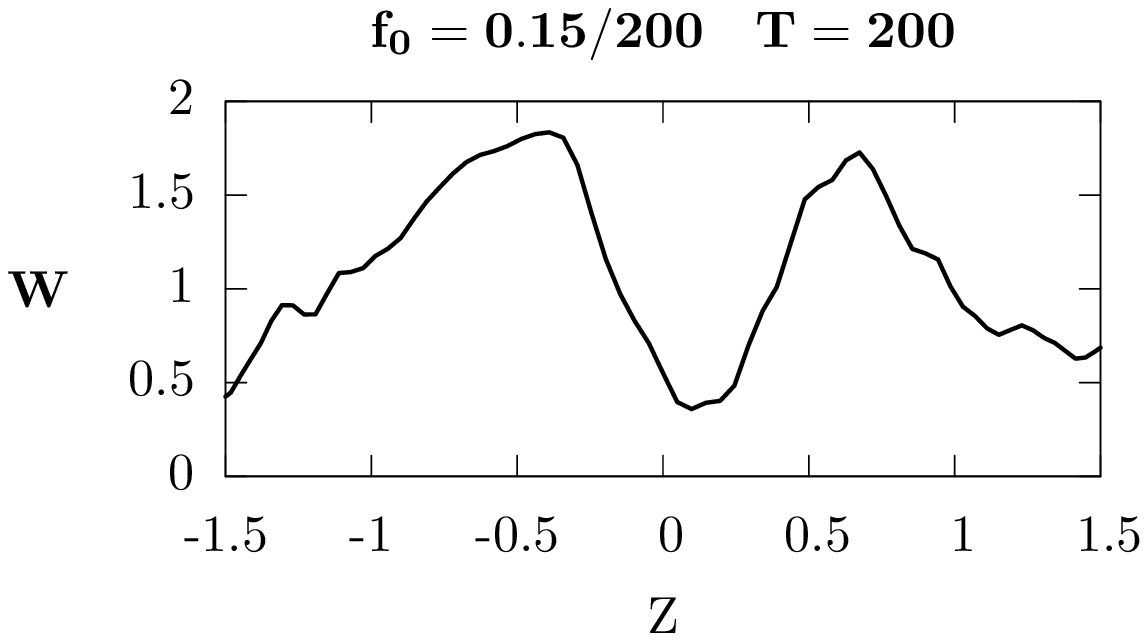}
	\includegraphics[width=0.45\textwidth]{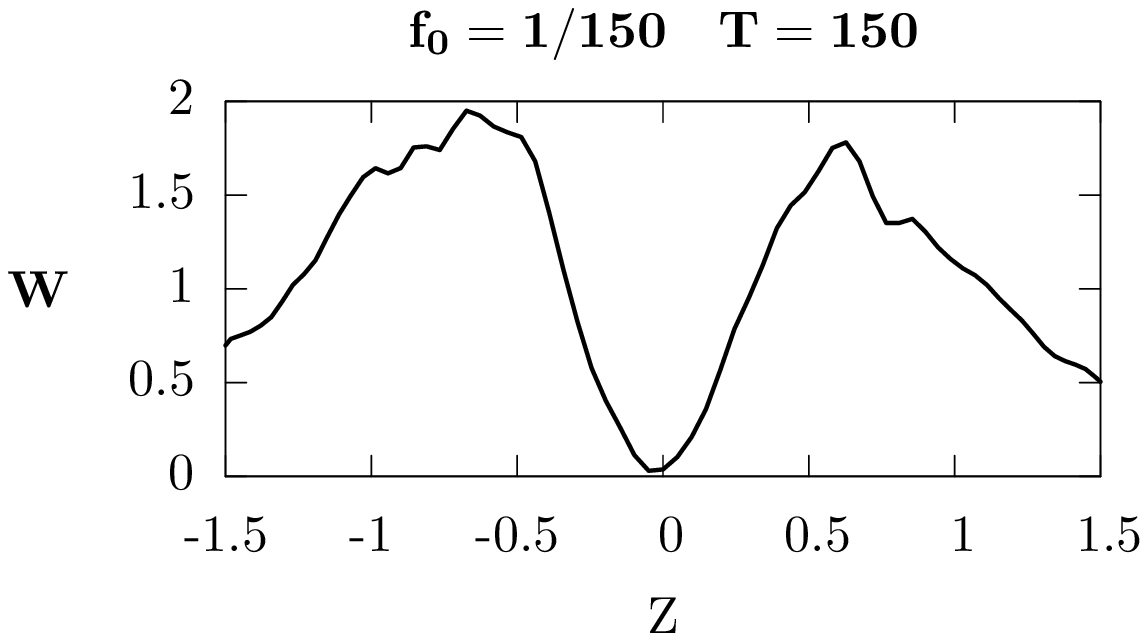}
	\includegraphics[width=0.45\textwidth]{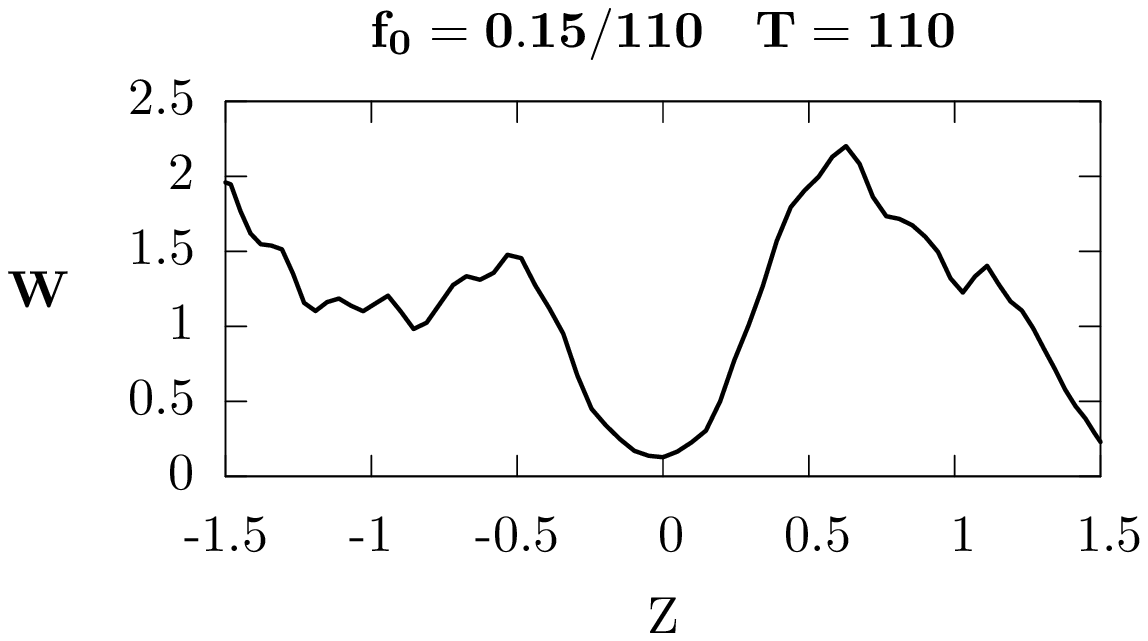}
	\includegraphics[width=0.45\textwidth]{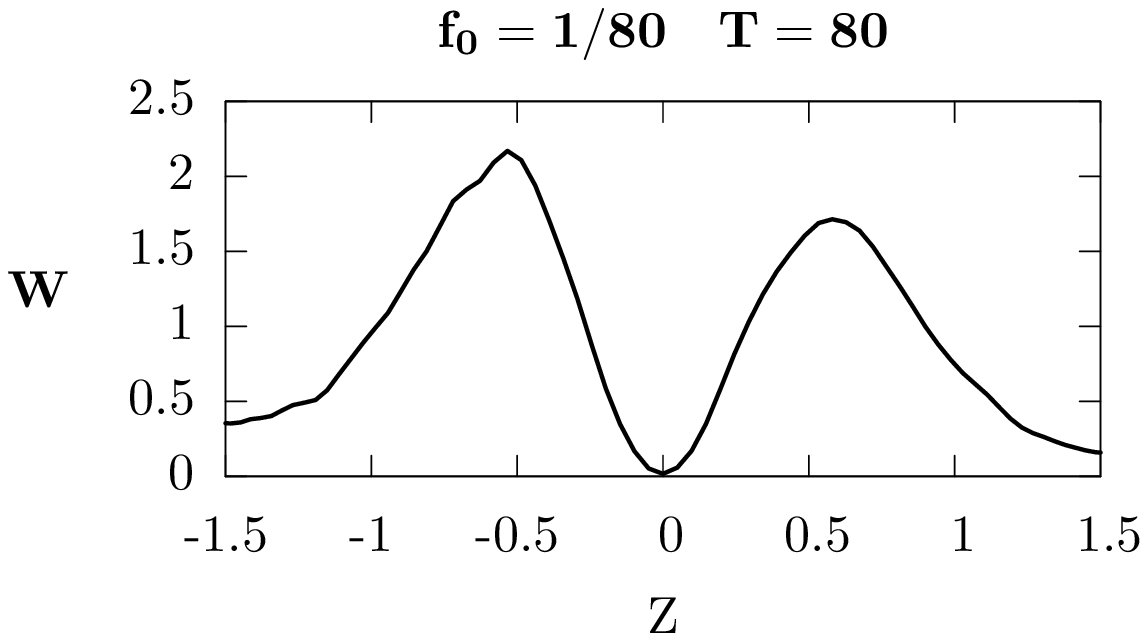}
	\includegraphics[width=0.45\textwidth]{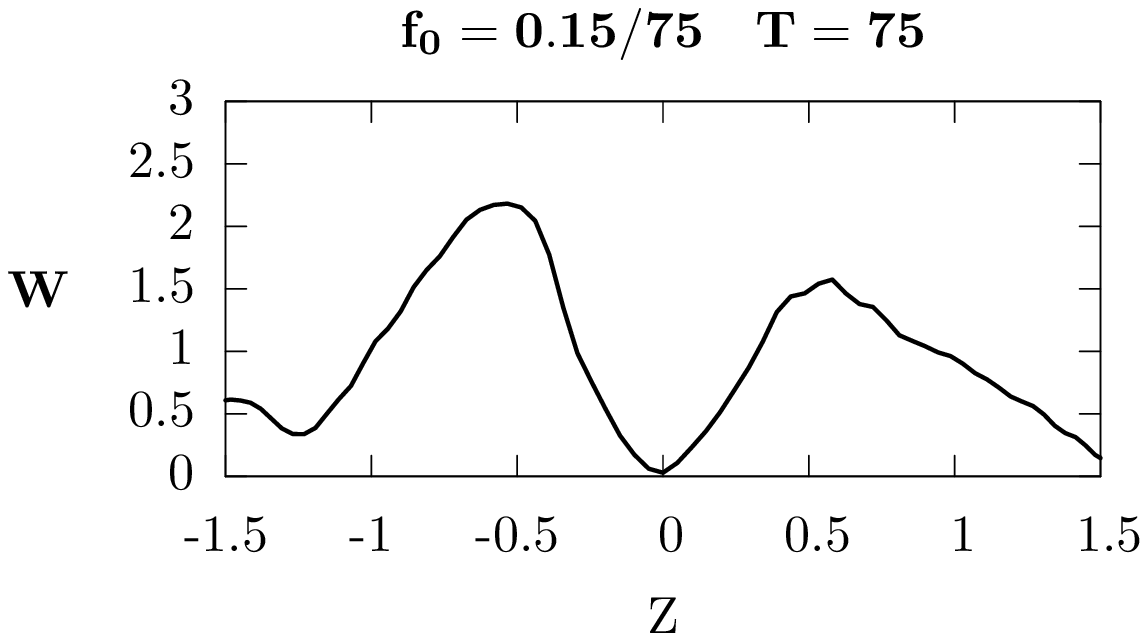}
	\includegraphics[width=0.45\textwidth]{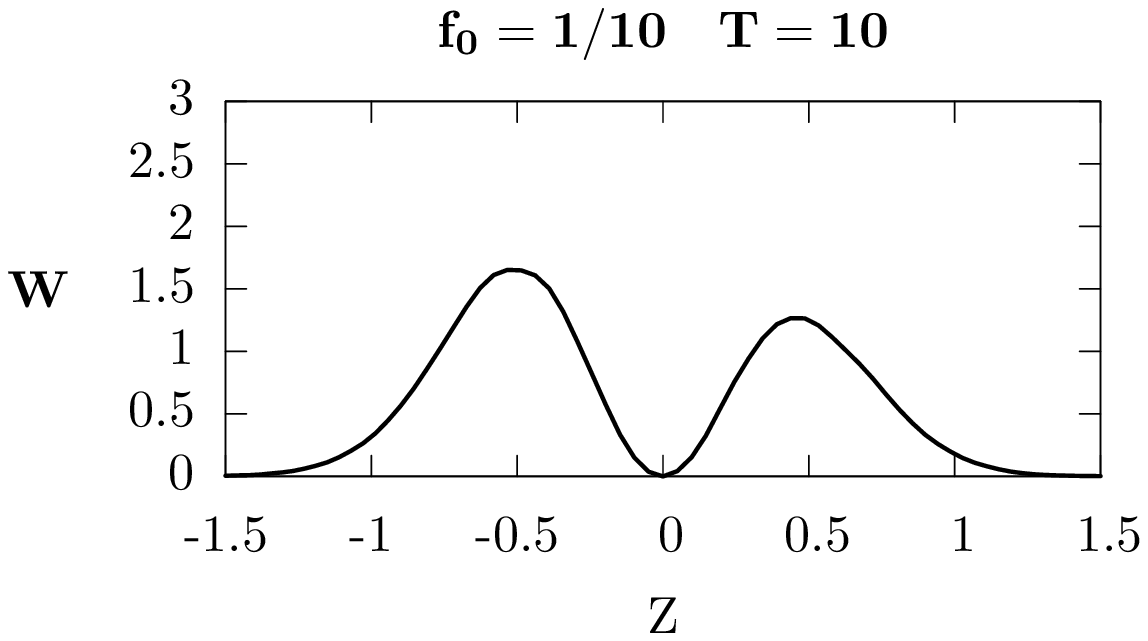}
	\includegraphics[width=0.45\textwidth]{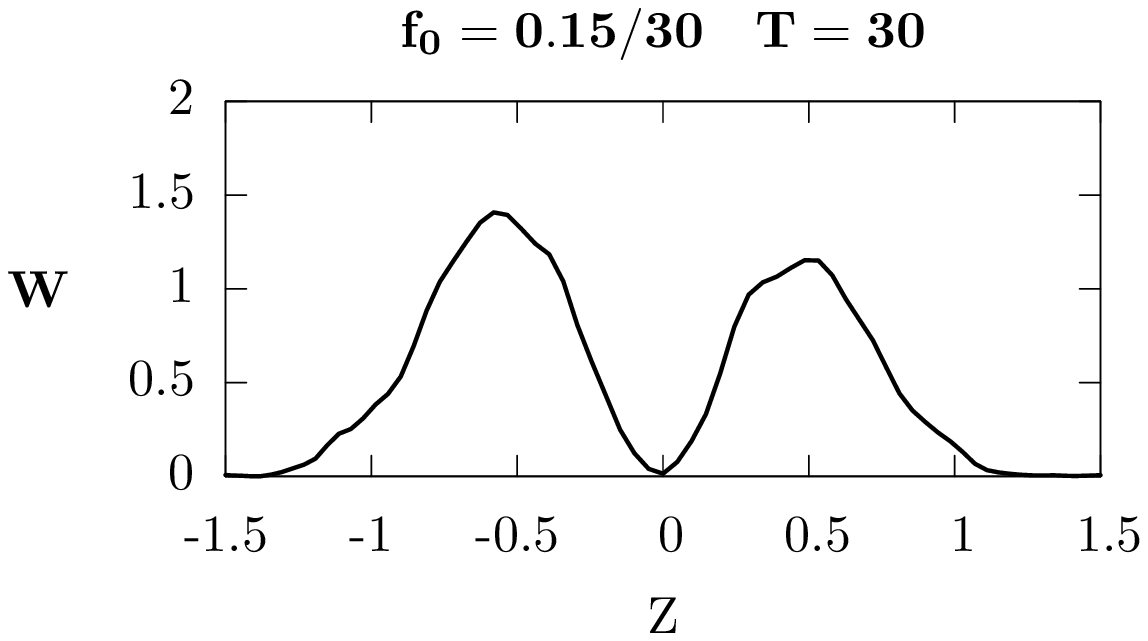}
	\caption{The energy dissipation rate due to the effective viscosity at
	each depth for the two sets of simulations with $z$ dependent $x$ 
	forcing with fixed amplitude and
	variable period. Left: strong forcing, right: weak forcing. The plots
	are for the same simulations as the plots in figure 
	\ref{fig: XZ_W_fit}. The same scaling has been applied to each plot as
	in figure \ref{fig: XZ_W_fit}}
	\label{fig: XZ_W_visc}
\end{center}
\end{figure}

\begin{figure}
\begin{center}
	\includegraphics[width=0.45\textwidth]{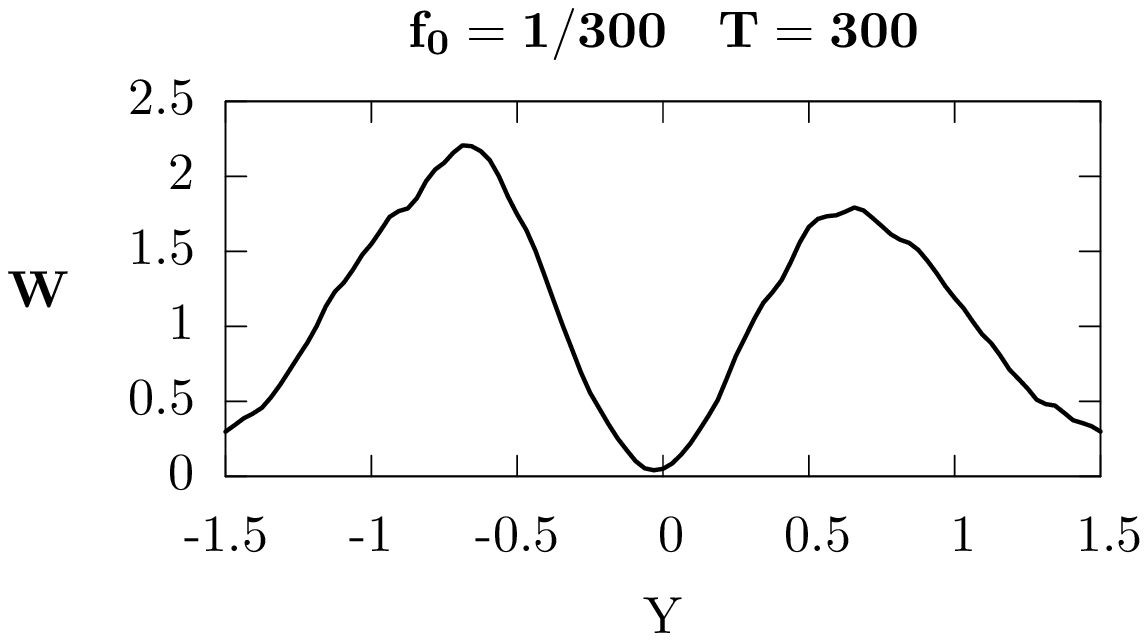}
	\includegraphics[width=0.45\textwidth]{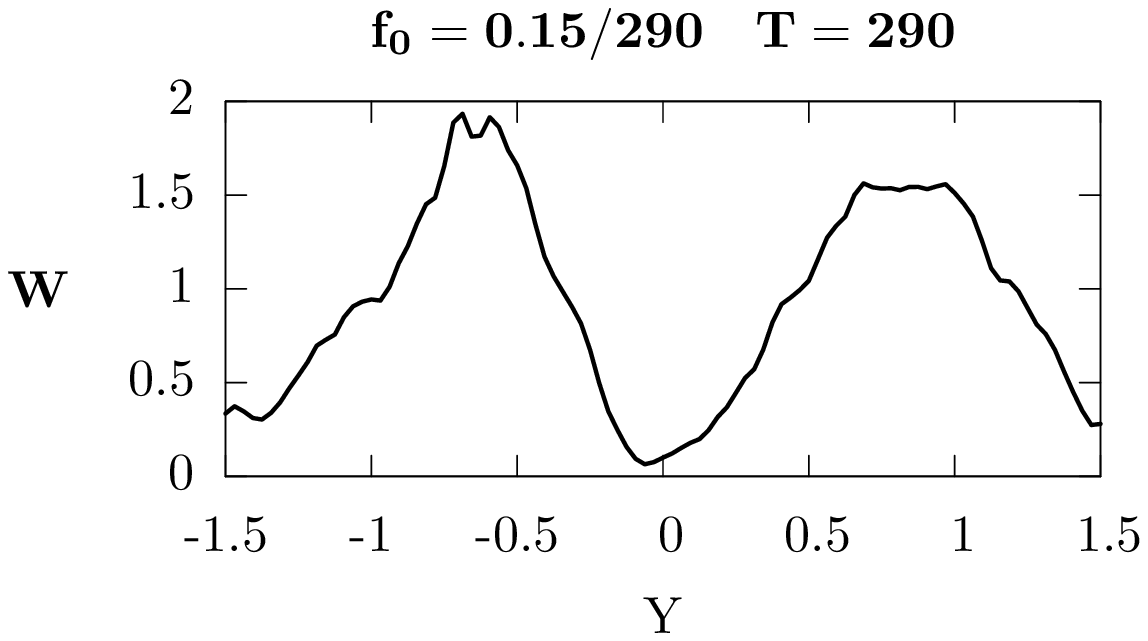}
	\includegraphics[width=0.45\textwidth]{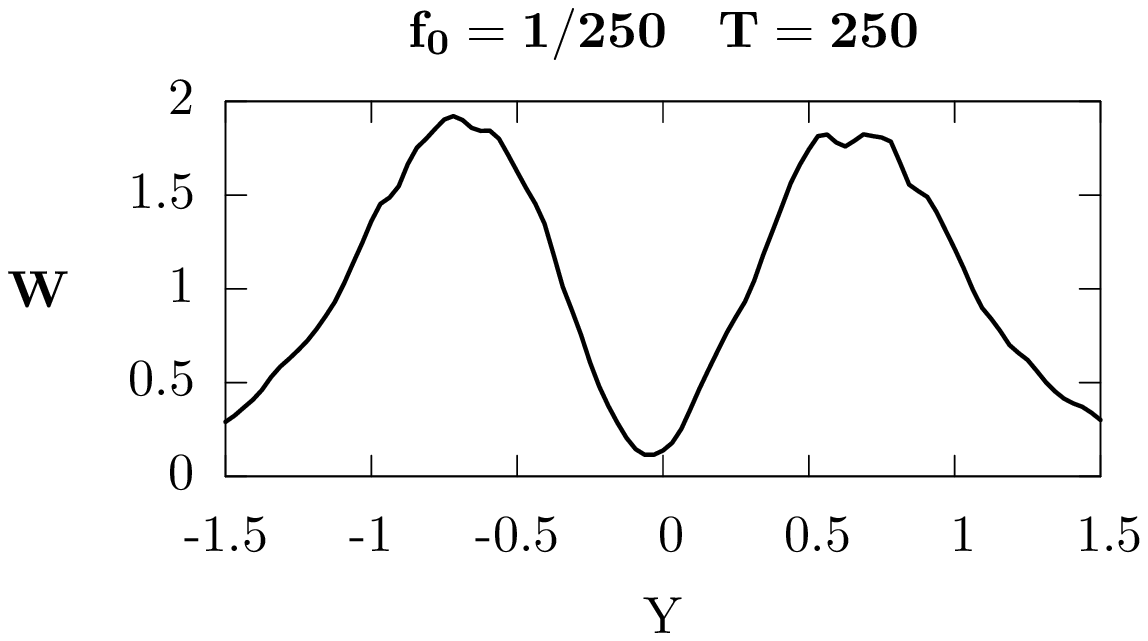}
	\includegraphics[width=0.45\textwidth]{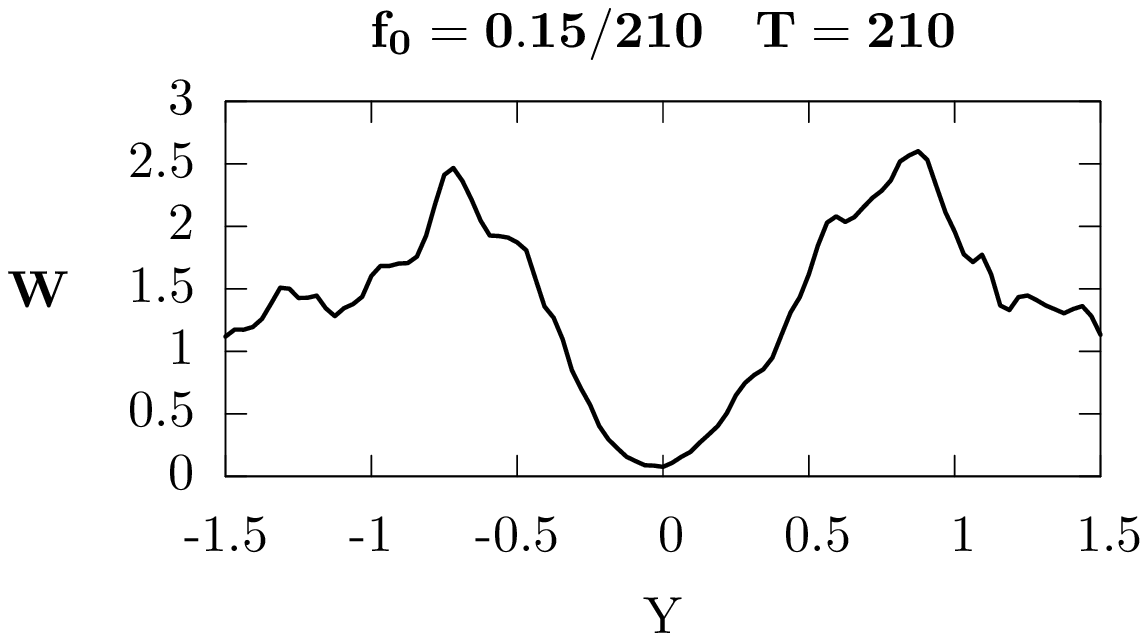}
	\includegraphics[width=0.45\textwidth]{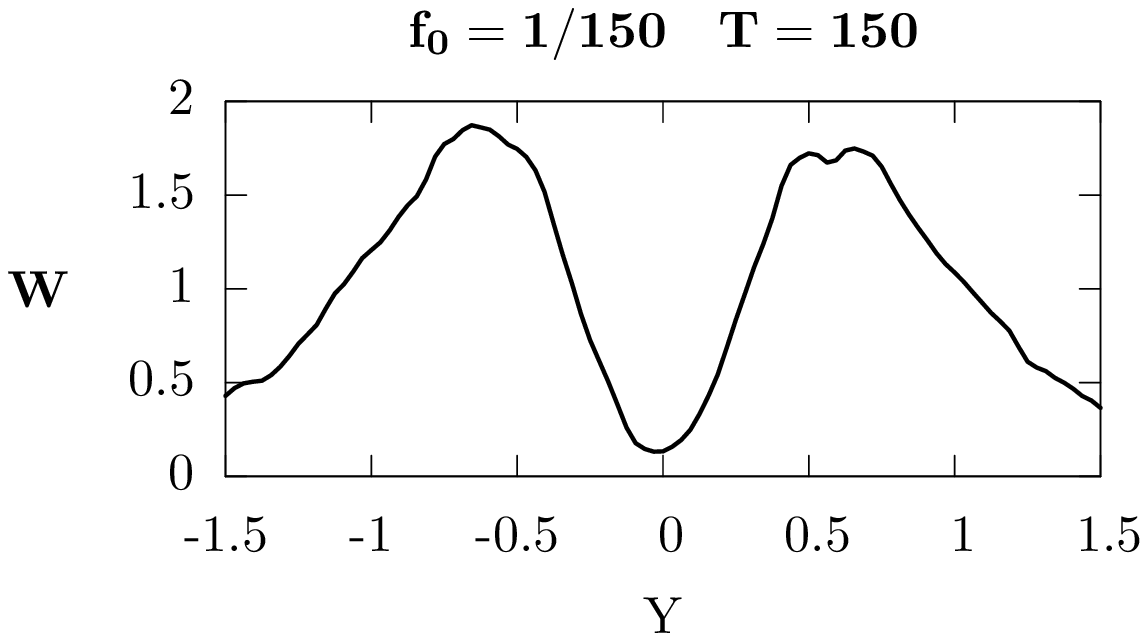}
	\includegraphics[width=0.45\textwidth]{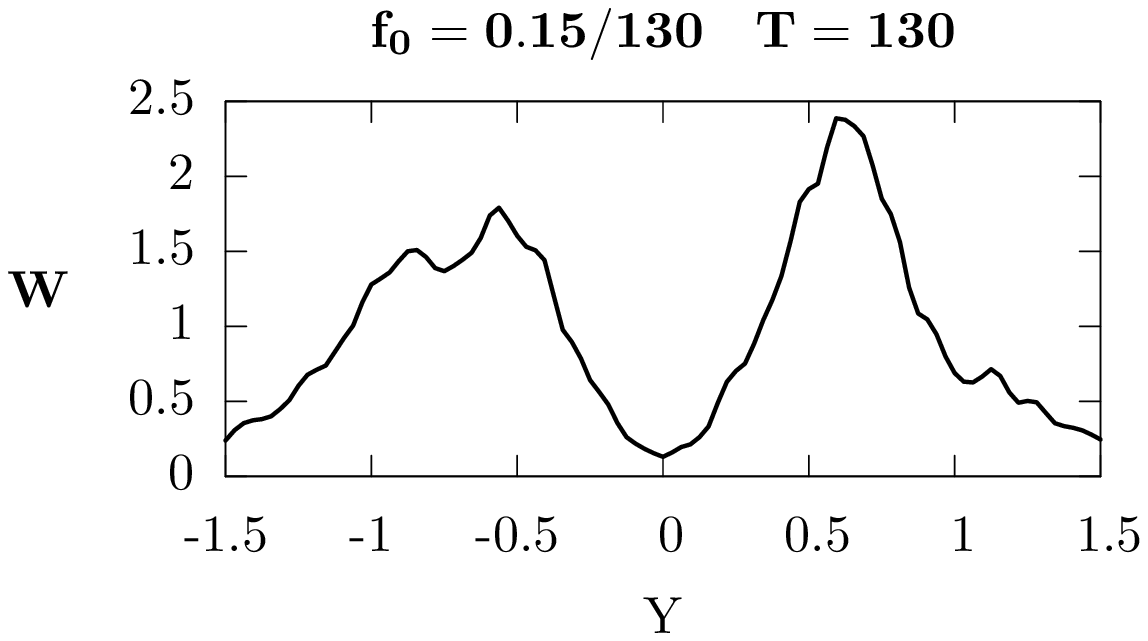}
	\includegraphics[width=0.45\textwidth]{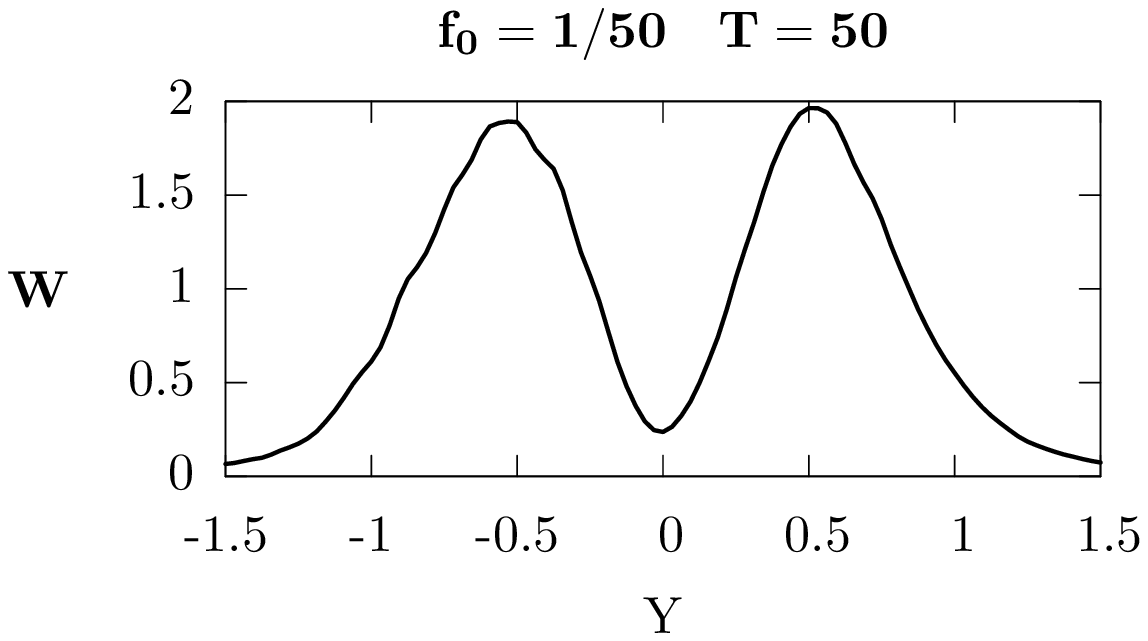}
	\includegraphics[width=0.45\textwidth]{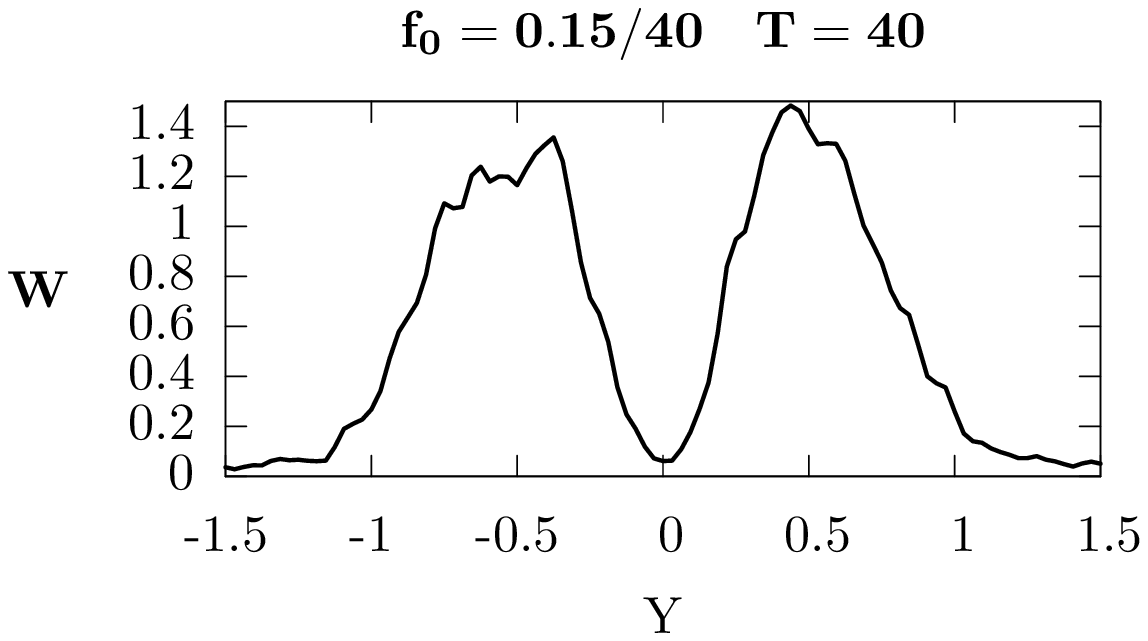}
	\caption{The energy dissipation rate due to the effective viscosity at
	each y plane for the two sets of simulations with $y$ dependent $x$ 
	forcing with fixed amplitude and variable period. Left: strong forcing,
	right: weak forcing. The plots are for the same simulations as the plots
	in figure \ref{fig: XY_W_fit}.The same scaling has been applied to each
	plot as in figure \ref{fig: XY_W_fit}}
	\label{fig: XY_W_visc}
\end{center}
\end{figure}

\begin{figure}
\begin{center}
	\includegraphics[width=0.45\textwidth]{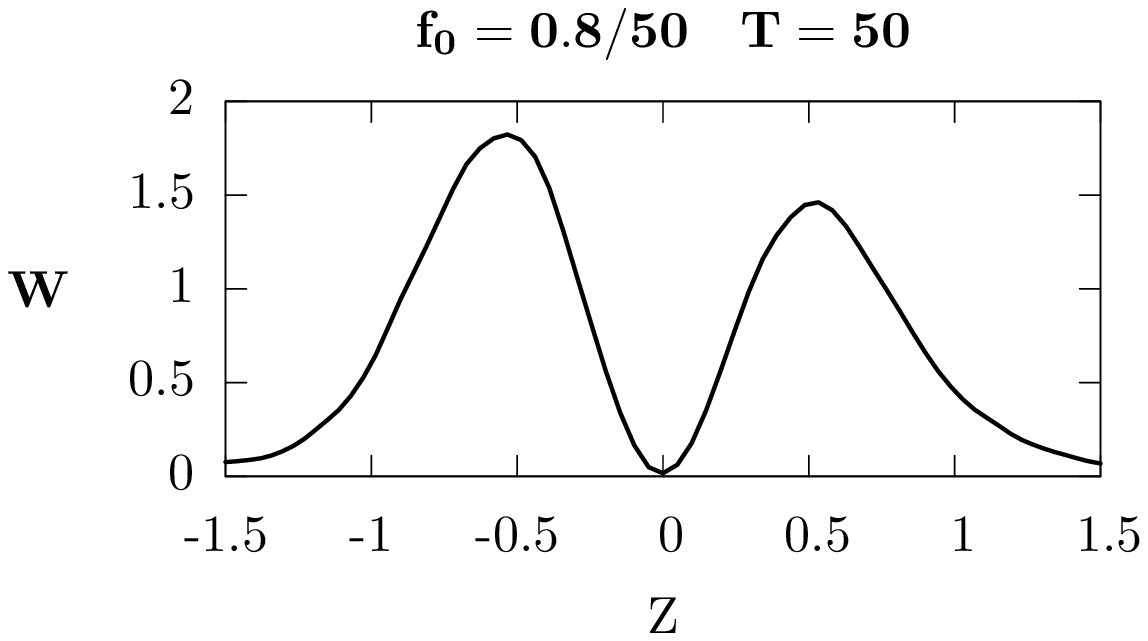}
	\includegraphics[width=0.45\textwidth]{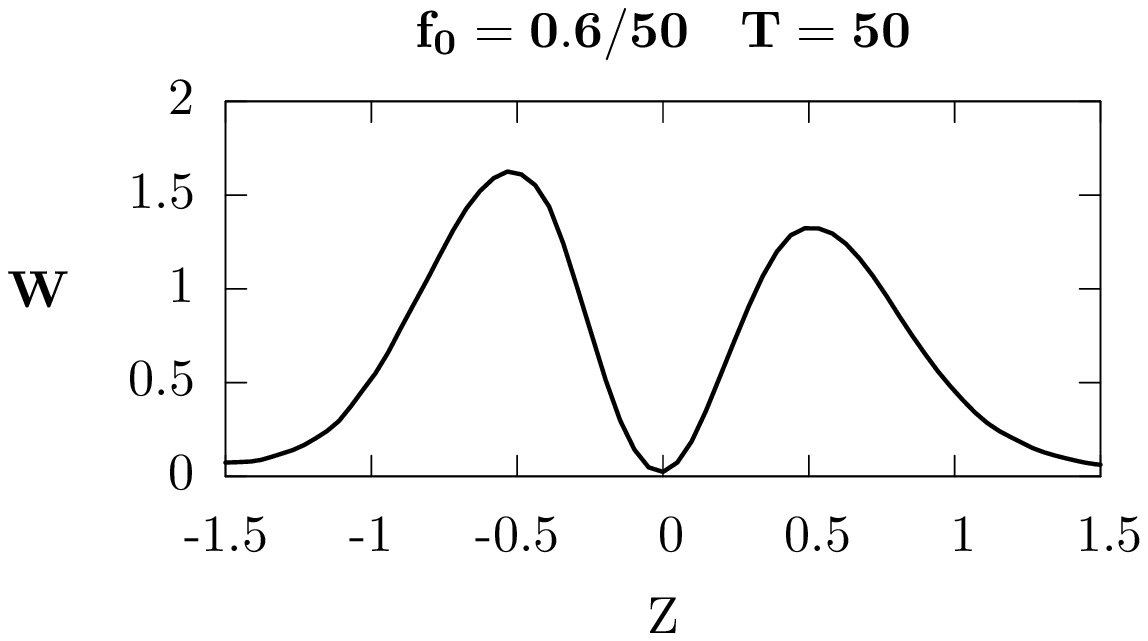}
	\includegraphics[width=0.45\textwidth]{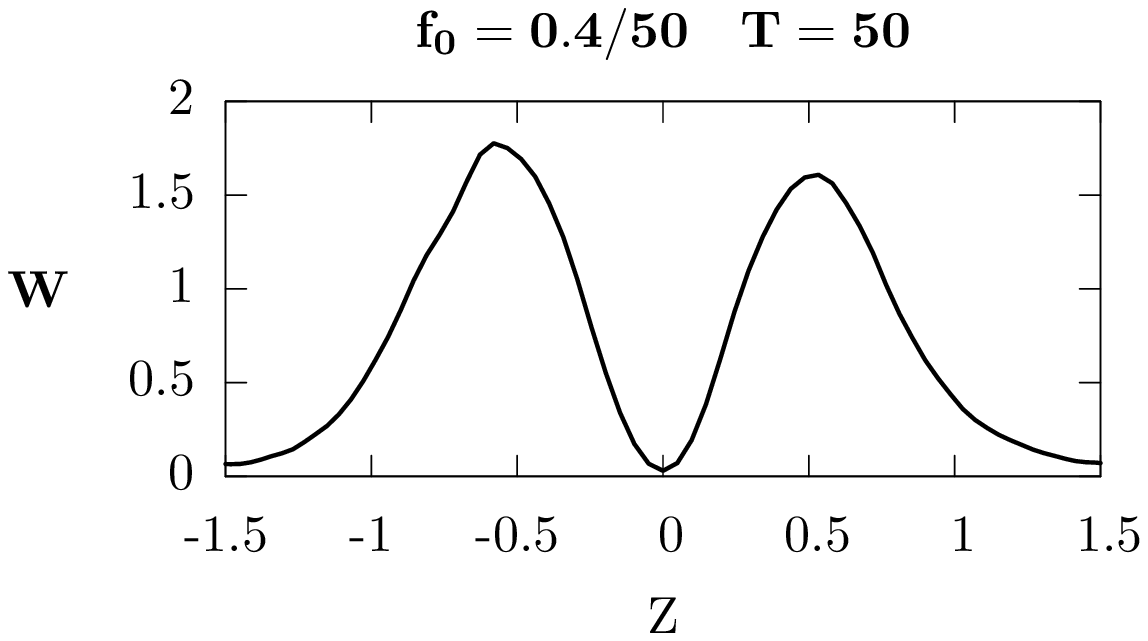}
	\includegraphics[width=0.45\textwidth]{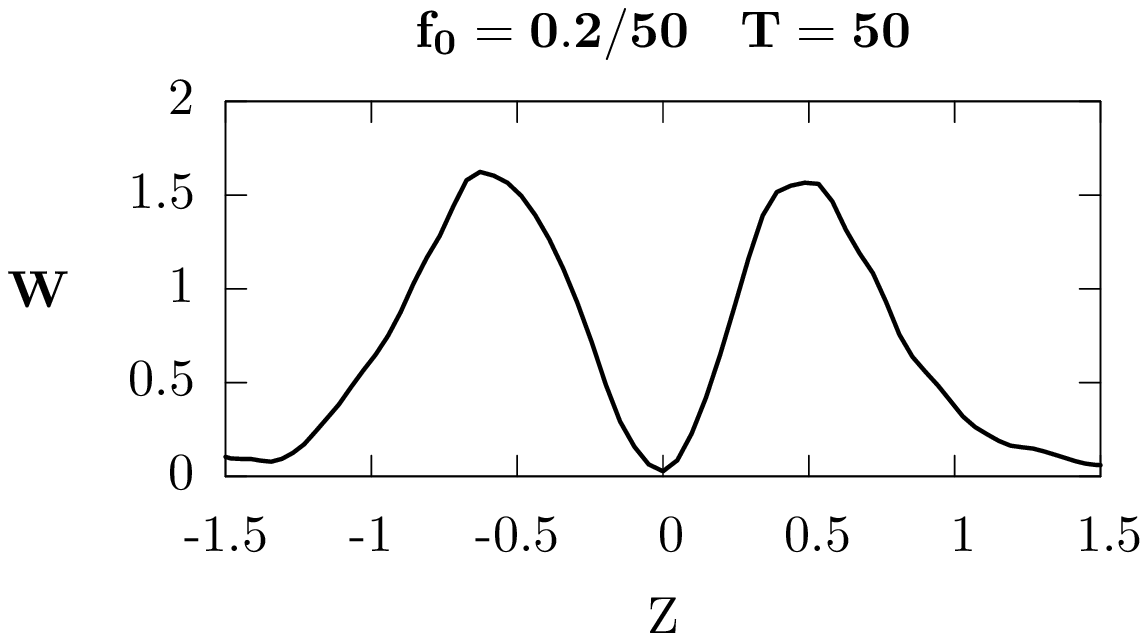}
	\caption{The energy dissipation rate due to the effective viscosity at
	each depth for the set of simulations with $z$ dependent $x$ forcing
	with fixed period and variable amplitude. The plots are for the same
	simulations as the plots in figure \ref{fig: Amp_W_fit}.The same scaling
	has been applied to each plot as in figure \ref{fig: Amp_W_fit}}
	\label{fig: Amp_W_visc}
\end{center}
\end{figure}

Plots of $\widetilde{W}_{xz}^{visc}(z)$ and
$\widetilde{W}_{xy}^{visc}(y)$ for the same cases as in figures \ref{fig:
XZ_W_fit}, \ref{fig: XY_W_fit} and \ref{fig: Amp_W_fit} are shown in figures 
\ref{fig: XZ_W_visc}, \ref{fig: XY_W_visc} and \ref{fig: Amp_W_visc}
respectively, where we have taken $z_{max}=-z_{min}=1.5$ for the depth
dependent forcing and $z_{max}=-z_{min}=1.8$ for the $y$ dependent forcing.

\begin{figure}
\begin{center}
	\includegraphics[width=0.8\textwidth,height=0.45\textwidth]{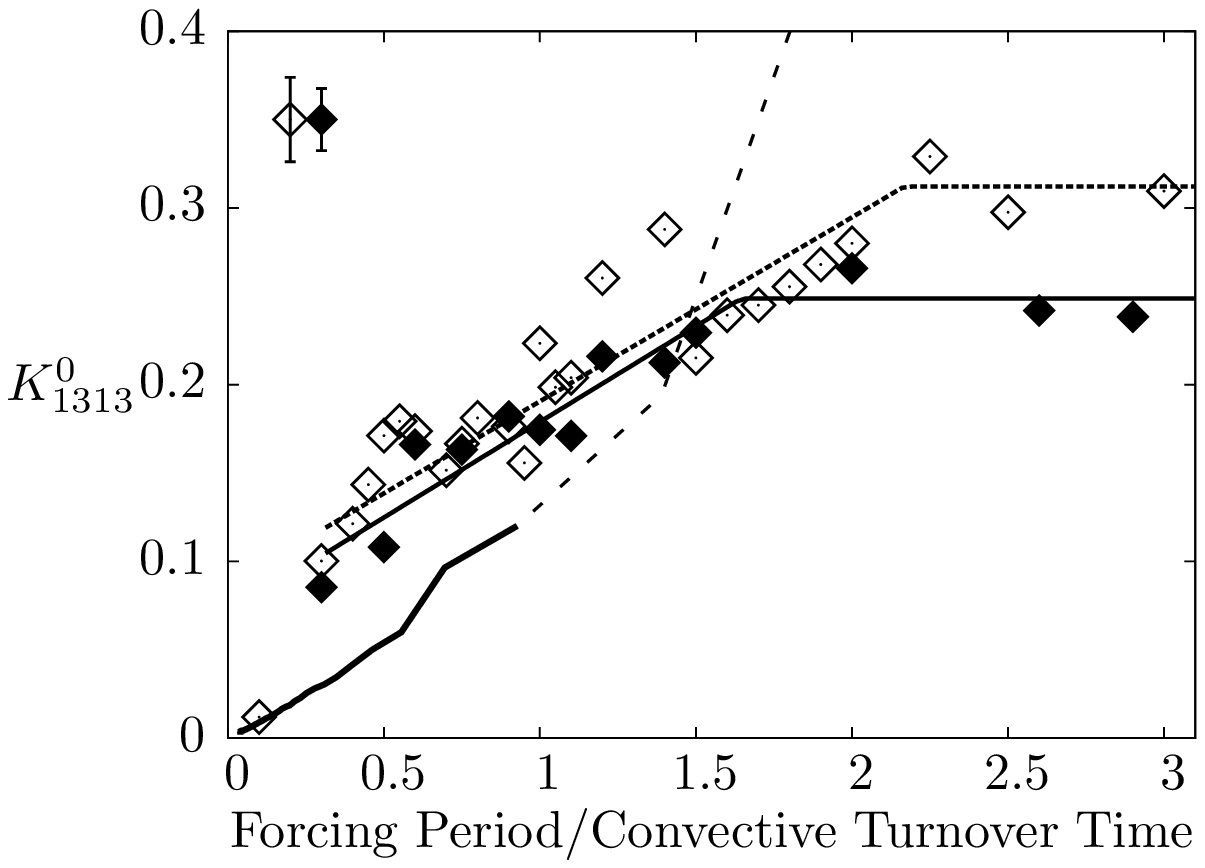}
	\includegraphics[width=0.8\textwidth,height=0.45\textwidth]{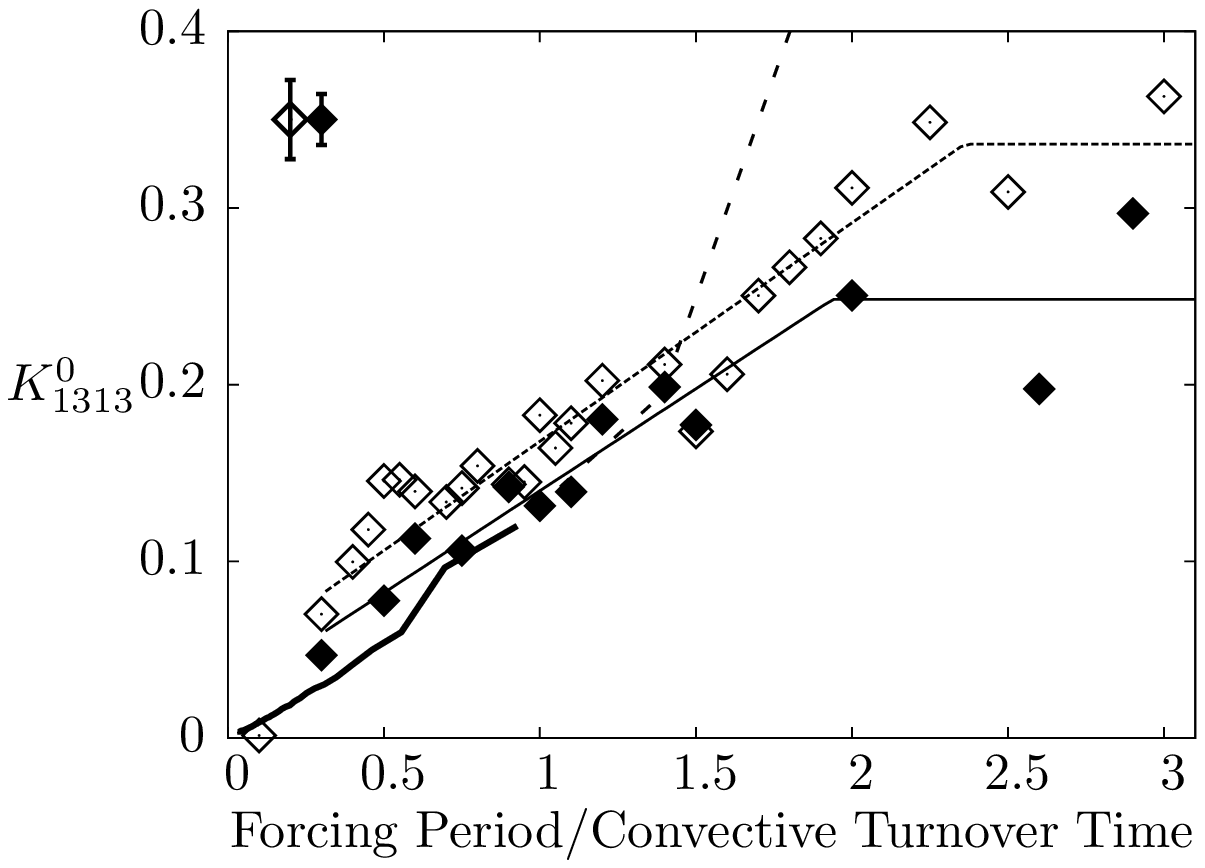}
	\caption{Comparison between the after-the-fact perturbatively estimated
	$K_{1313}^0$ (curve) and the viscosity obtained from the direct
	simulations by least squares fit of $W_{xz}^{visc}$ 
	(equation \ref{eq: Wxz_visc}) to $W_{xz}^{turb}$  (equation \ref{eq: Wxz_turb}) 
	--- top ; and from setting $\int
	W_{xz}^{turb}(z)=\int\widetilde{W}_{xz}^{visc}(z)$ --- bottom. 
	The strong forcing points are
	plotted with empty symbols, and the weak forcing with filled symbols. 
	The horizontal axis is the perturbation period ($T$) in units of
	the convective turnover time in the box.
	Also shown are least squares fits to the strong forcing points (dashed
	line) and the weak forcing points (solid line), by a linear function
	with saturation. The error bars in the top left corner of the  plots
	correspond to the standard deviation of $K_{1313}^0$ (assumed the same
	for all points) obtained from the differences with the fitted curves.}
	\label{fig: xz_visc_comparison}
\end{center}
\end{figure}

\begin{figure}
\begin{center}
	\includegraphics[width=0.8\textwidth,height=0.45\textwidth]{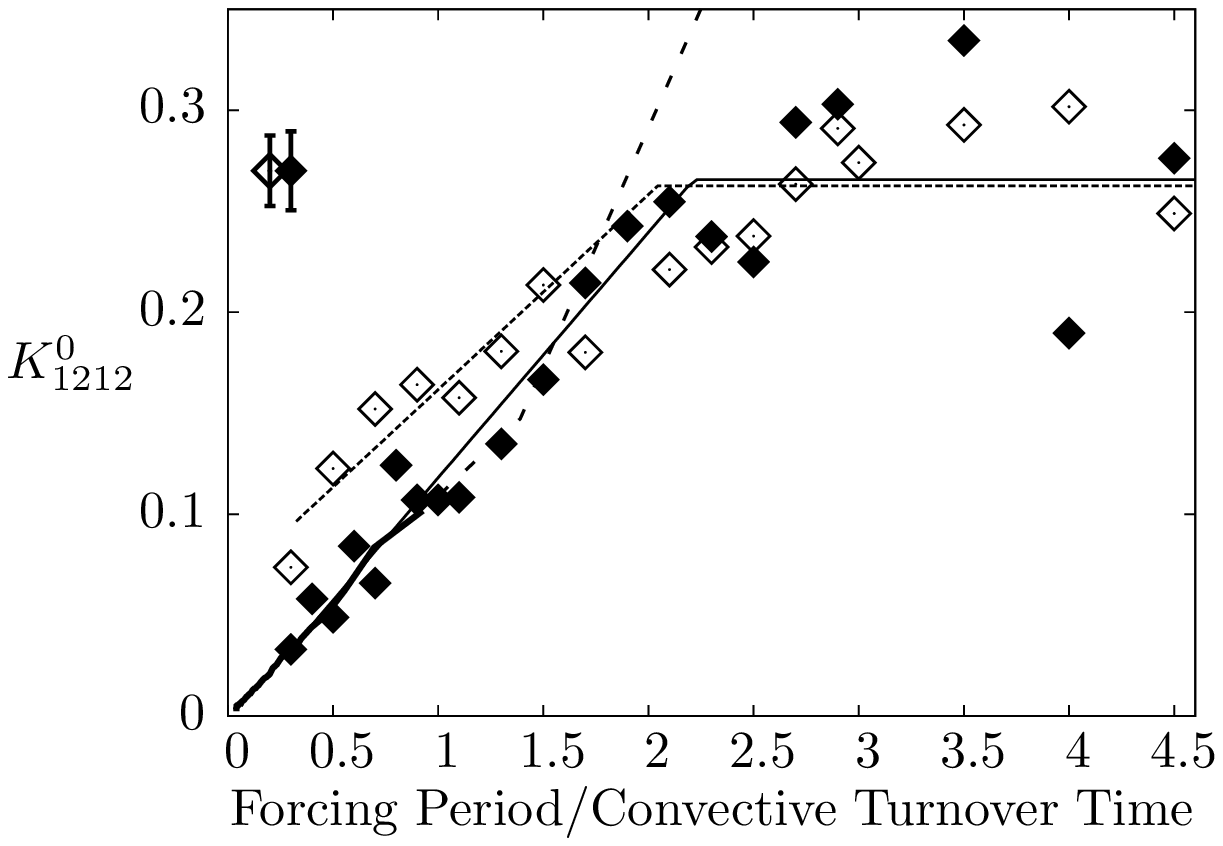}
	\includegraphics[width=0.8\textwidth,height=0.45\textwidth]{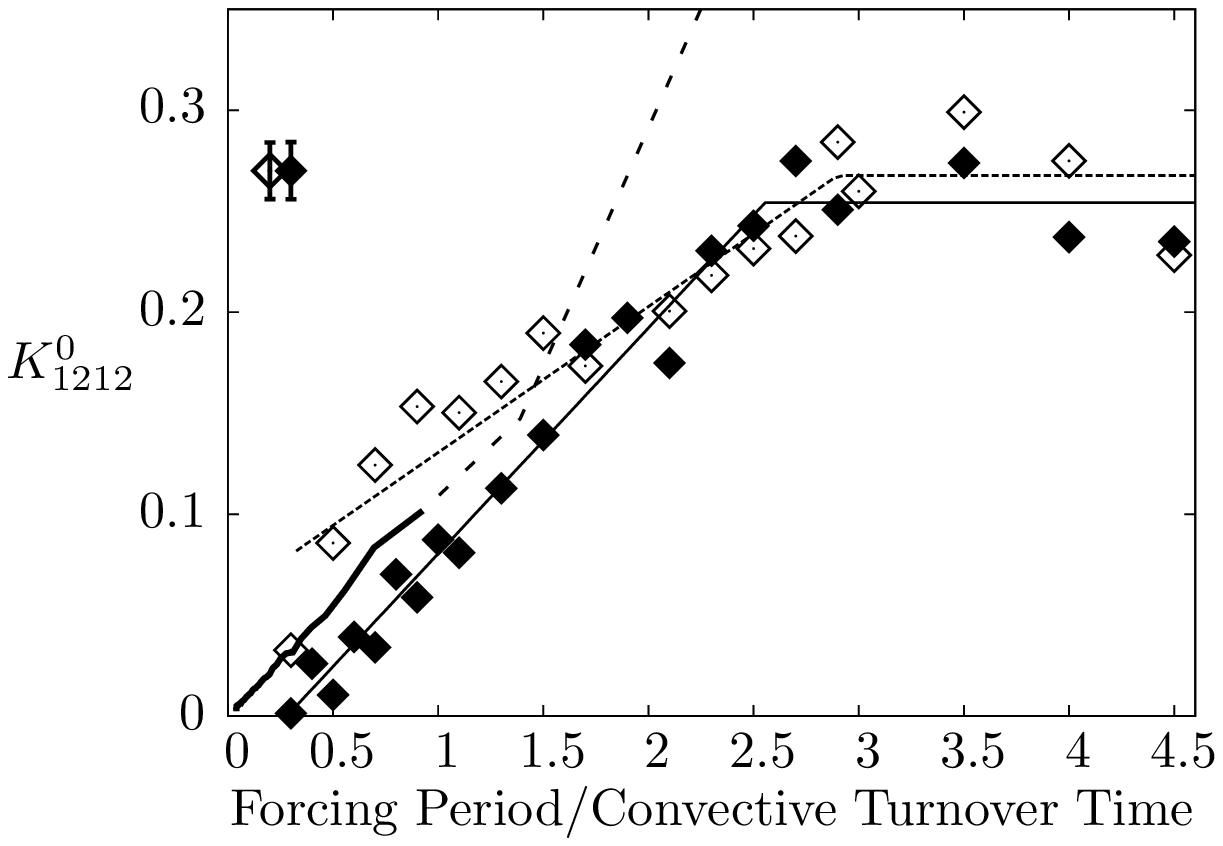}
	\caption{Comparison between the after-the-fact perturbatively estimated
	$K_{1212}^0$ (curve) and the viscosity obtained from the direct
	simulations by least squares fit of $W_{xy}^{visc}$ 
	(equation \ref{eq: Wxy_visc}) to $W_{xy}^{turb}$  (equation \ref{eq: Wxy_turb}) 
	--- top ; and from setting 
	$\int W_{xy}^{turb}(y)=\int\widetilde{W}_{xy}^{visc}(y)$ --- bottom. 
	The strong forcing points are
	plotted with empty symbols, and the weak forcing with filled symbols. 
	The horizontal axis is the perturbation period ($T$) in units of
	the convective turnover time in the box. Also shown are least
	squares fits to the strong forcing points (dashed line) and the weak
	forcing points (solid line) by a linear function with saturation.
	The error bars in the top left corner of the  plots correspond to the standard
	deviation of $K_{1212}^0$ (assumed the same for all points) obtained from 
	the differences with the fitted curves.}
	\label{fig: xy_visc_comparison}
\end{center}
\end{figure}

\subsection{Comparison Between the Perturbative and Direct Calculation}
\label{sec: comparison}

We compare the \citet{Penev_Barranco_Sasselov_08a} perturbative estimate of the
$K_{1313}$ and $K_{1212}$ viscosity coefficients to
the directly calculated effective viscosity, obtained by the procedures of
section \ref{sec: direct dissipation}, in figures \ref{fig: xz_visc_comparison}
and \ref{fig: xy_visc_comparison}.

The curves in those figures correspond to the perturbative expansion, and the
points correspond to the direct calculation. Since the
perturbative calculation assumes that the forcing period is small compared to
the convective turnover time, we have used a solid line only for $T<\tau_c$.
For longer periods (dotted line) this method is certainly not applicable.

We see that the effective viscosity predicted with all three methods scales
linearly with the forcing period for most of the range covered by our
simulations. Further, the direct calculations  show that the effective viscosity
saturates for long periods, after which it remains roughly constant. 
One physically
expects to see this saturation, because
for forcing periods much larger than any convective timescale there is no
reason why the dissipation efficiency of the convective zone should depend on
the period. The exact saturation period varies between the different forcing
cases and even viscosity estimation methods, but it is seen to be somewhere in
the range $1.5\tau_c<T_{sat}<3\tau_c$.

In order to obtain a functional dependence for the effective viscosity on period
we perform a least square fit to the direct calculation points, where we fit a
linear function that saturates at some period. The resulting fits
are shown in figures \ref{fig: xz_visc_comparison} and \ref{fig:
xy_visc_comparison} as solid lines for the weak forcing cases and dashed lines
for the strong forcing cases. The parameters of the fitted lines are given in
tables  \ref{tbl: xz visc fit params} and \ref{tbl: xy visc fit params}. 
This also allows us to get an estimate of the error bars associated
with the points. Those are shown in the upper left corners of the plots in
figures \ref{fig: xz_visc_comparison} and \ref{fig: xy_visc_comparison}.

\begin{table}[tb]
\begin{center}
	\caption{The linear regression parameters corresponding to the solid and
	dashed lines in fig. \ref{fig: xz_visc_comparison}.}
	\label{tbl: xz visc fit params}

	\begin{tabular}{l c c|c c}
		& \multicolumn{2}{c|}{Depth Fit}
		& \multicolumn{2}{|c}{Dissipation Match}
		\\
				& strong forcing& weak forcing	
				& strong forcing& weak forcing\\
		\hline
		slope			& $0.10\pm0.012$& $0.11\pm0.015$ 
					& $0.12\pm0.009$& $0.11\pm0.02$\\
		zero crossing		& $0.086\pm0.013$& $0.071\pm0.014$
					& $0.044\pm0.011$& $0.02\pm0.02$\\
		saturation period	& $2.2\pm0.2$& $1.64\pm0.15$ 
					& $2.4\pm0.2$& $2.0\pm0.2$
	\end{tabular}
\end{center}
\end{table}

\begin{table}[tb]
\begin{center}
	\caption{The linear regression parameters corresponding to the solid and
	dashed lines in fig. \ref{fig: xy_visc_comparison}.}
	\label{tbl: xy visc fit params}

	\begin{tabular}{l c c|c c}
		& \multicolumn{2}{c|}{Depth Fit}
		& \multicolumn{2}{|c}{Dissipation Match}
		\\
				& strong forcing& weak forcing	
				& strong forcing& weak forcing\\
		\hline
		slope			& $0.097\pm0.023$& $0.12\pm0.02$ 
					& $0.073\pm0.023$& $0.11\pm0.01$\\
		zero crossing		& $0.065\pm0.023$& $-0.003\pm0.023$
					& $0.058\pm0.014$& $-0.03\pm0.008$\\
		saturation period	& $2.0\pm0.3$& $2.2\pm0.2$ 
					& $2.9\pm0.2$& $2.6\pm0.1$
	\end{tabular}
\end{center}
\end{table}

Since for very short periods the perturbative calculation should be valid, and
the turbulence should be well approximated by a Kolmogorov cascade, we expect
that the scaling of the effective viscosity with period should be quadratic for
those short periods. The proportionality constant for this steeper scaling
should determine the zero crossings of the best fit linear approximations to the
effective viscosity. From tables
\ref{tbl: xz visc fit params} and \ref{tbl: xy visc fit params} we see that
these are mostly positive, except for $K_{1212}^0$
in the weak forcing case, where for the fitting method the zero crossing is
basically indistinguishable from zero, and in the dissipation matching method it
is slightly negative.

We believe that this negative zero crossing is caused by
the fact that the forcing in this case is not small near the boundaries, and as
a result significant energy is deposited there by the external forcing,
especially since in those regions the horizontal velocity is very large, due to
the collision of the mainly vertical flow in the bulk of the box with the
impenetrable boundaries.
This large horizontal flow is non physical and as such we do not include the
regions near the boundaries in our calculations. Ignoring this external energy
source is not important for the fitting of the spatial dependence of the
deposited energy since it will be ignored from both $W_{xy}^{turb}$ and
$W_{xy}^{visc}$. However, if some amount of this energy makes it to the region
of the box used to calculate $\widetilde{W}_{xy}^{visc}$ before the turbulent
cascade has dissipated it, it will act to artificially increase
$\widetilde{W}_{xy}^{visc}$, and consequently decrease the effective viscosity
we calculate.

\subsubsection{Amplitude Dependence}
The three dependences in each of the plots of figures  \ref{fig:
xz_visc_comparison} and \ref{fig: xy_visc_comparison} correspond to three
different forcing strengths: strong forcing with $v_{forc}/v_{conv}\approx2.7$,
weak forcing with $v_{forc}/v_{conv}\approx0.4$ and the pertubative calculation
with $v_{forc}/v_{conv}\ll1$, where $v_{forc}$ is the peak velocity due to the
external forcing and $v_{conv}$ is the root mean square velocity for the central
plane of the box in the absence of forcing.

\begin{figure}[tb]
\begin{center}
	\includegraphics[width=0.45\textwidth]{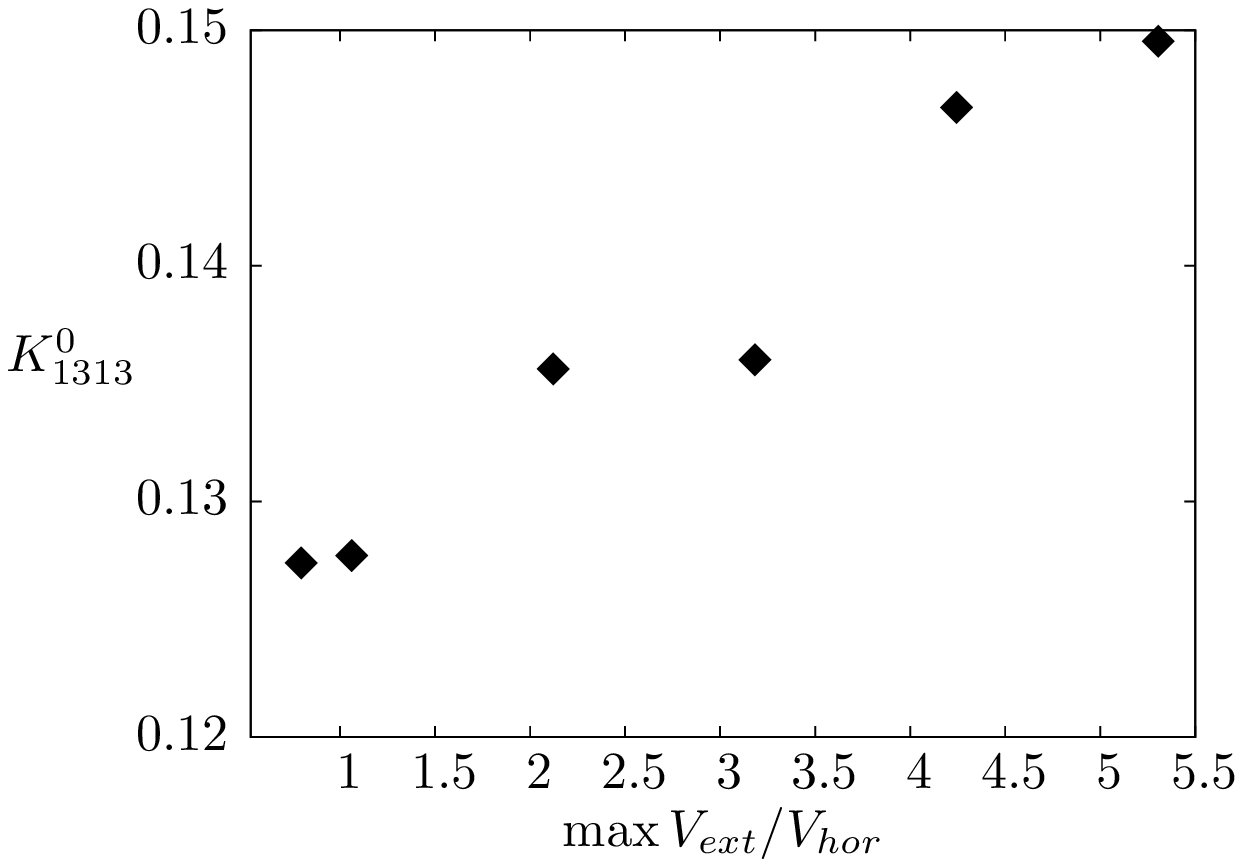}
	\includegraphics[width=0.45\textwidth]{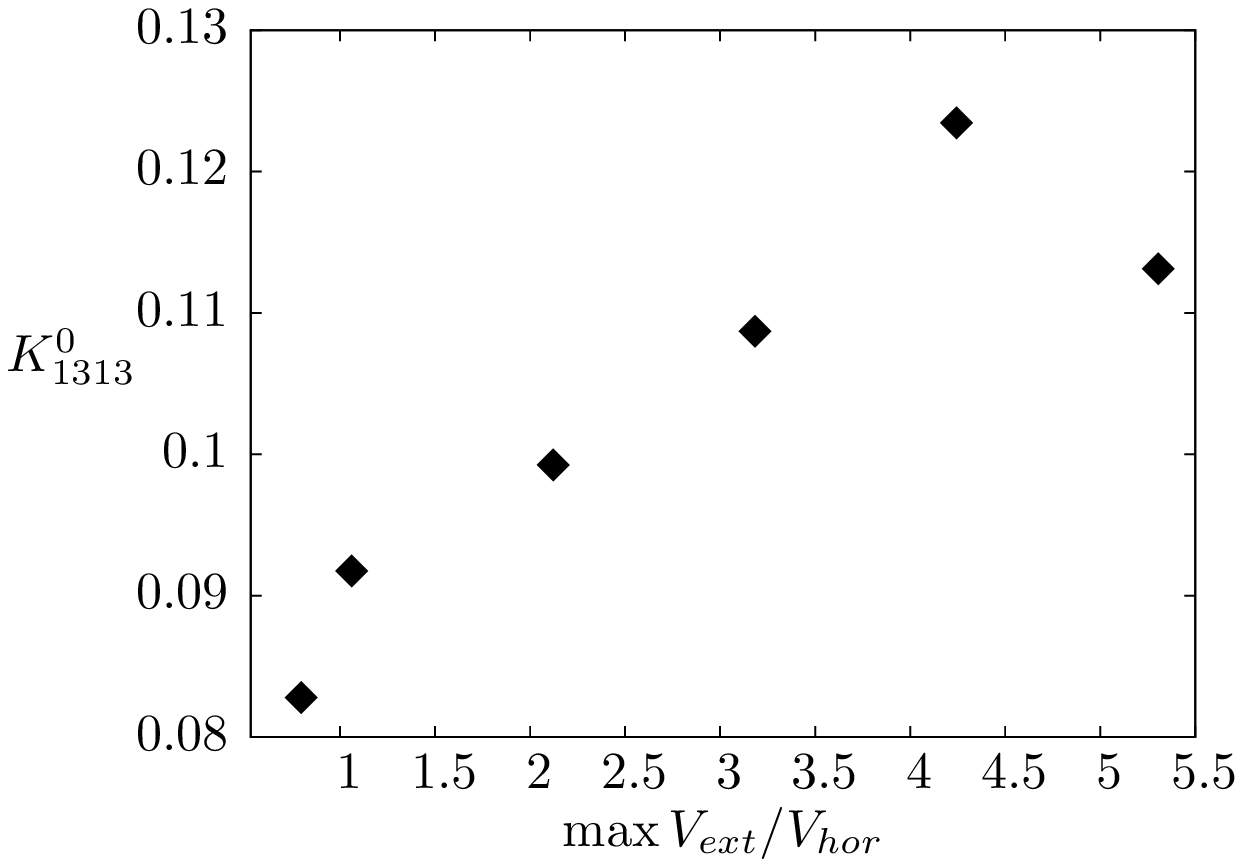}
	\caption{The dependence of the effective viscosity $K_{1313}^0$ on the 
	forcing strength ($\max V_{ext}/V_{hor}$), where $V_{hor}$ is the
	horizontal r.m.s. velocity in the absence of forcing. The left plot
	correspond to the effective viscosity
	estimated by least squares fitting of $W_{xz}^{visc}$ 
	(equation \ref{eq: Wxz_visc}) to $W_{xz}^{turb}$  (equation \ref{eq: Wxz_turb})
	and the right plot corresponds to the effective viscosity obtained by 
	setting $\int W_{xz}^{turb}(z)=\int\widetilde{W}_{xz}^{visc}(z)$}
	\label{fig: A dependence}
\end{center}
\end{figure}

The systematic difference between the three viscosities suggests a possibly
important amplitude dependence of the dissipation efficiency. In order to get
some idea for the importance of the magnitude of the external shear in
determining the values of $K_{1313}^0$ we performed four additional simulations
with $z$ dependent external forcing with period $T=\tau_c/2$ with strengths
intermediate between the
strong and weak forcing cases considered above (see table \ref{tbl:amp runs}
for the details of each run and figures \ref{fig: Amp_W_fit} and \ref{fig:
Amp_W_visc} for the energy rate curves). Using the same two methods discussed
above, we estimate the effective viscosity for those cases, which we plot in
figure \ref{fig: A dependence}, where we have also added the value of the fitted
straight lines for the strong and weak forcing from figure \ref{fig:
xz_visc_comparison} at half the convective turnover time.

Clearly the effective viscosity increases for larger external forcing, but the
effect is much smaller than the period dependence. Further, the
perturbative calculation, which would correspond to a forcing strength of
zero has a value at $T=\tau_c/2$ of 0.055, which is far below the range of direct
calculation values obtained from either method. This is possibly due to the fact
that at the periods covered by our direct calculation points the value of
$T/\tau_c$ is not small enough for the perturbative calculation to be a good
approximation. Another possibility is that for weaker forcing the amplitude
dependence is much stronger, but we must note that according to the
perturbative calculation, for very small amplitudes, the effective viscosity is
independent of the amplitude of the external shear. 

From figure \ref{fig: xy_visc_comparison} we see that this discrepancy between
the direct calculation and the perturabative calculation does not exist for
$K_{1212}^0$ and the weak forcing direct calculation points lie close to the
the perturbative calculation curve.

A possible explanation for the observed amplitude dependence is that, at the
forcing amplitudes we have used, the external shear is itself going to drive
turbulence and lead to its own dissipation. Thus, the stronger the forcing, the
more additional dissipation is expected to occur. 

\section{Conclusions}
\label{sec: conclusions}
We have completed a set of simulations of turbulent convection relevant to
stellar surface convective zones, with external
forcing introduced directly in the momentum equation in the form of a periodic,
horizontal, position dependent gravitational acceleration. We then found an
effective viscosity assumed to be of the form of equations \ref{eq: nu_xz_form}
and \ref{eq: nu_xy_form} (in accordance with mixing length theory), using
two methods:
\begin{enumerate}
	\item{least squares fitting of the work done by the external forcing on
	the flow at each plane of constant forcing to the energy
	transported or dissipated away from that plane by the effective
	viscosity.}
	\item{matching the overall energy deposited into the box by the external
	forcing to the energy dissipated by the effective viscosity.}
\end{enumerate}
Both methods produce effective viscosity that scales linearly with period with
the same slope, but they differ by a constant offset. This offset is due to the
fact that, for most simulations, the negative side lobes of the energy
removed by viscosity are less deep relative to the positive central peak than
for the external work profile. For the least squares fitting method
this tends to drive the dimensionless proportionality constant to higher values
slightly over--predicting the central peak, but getting a bit deeper sidelobes.
For the energy matching method, this
tends to give a lower value for the viscosity in order to match the overall
integrals under the curves.

This seems to suggest that the particular depth dependence we assumed for the
turbulent viscosity might be slightly off. In particular it would seem to
require a bit larger values at the depths of these side lobes. 

Aside from this small effect, we
found that the one parameter fits of the first method were able to capture
the details of the observed position dependence of the deposited energy 
(see fig. \ref{fig: XZ_W_visc},
\ref{fig: XY_W_visc} and \ref{fig: Amp_W_visc}), which suggests that the
effective viscosity assumption for treating turbulent dissipation in convective
zones is valid.

We compared this directly obtained effective viscosity with the lowest
order \citet{Goodman_Oh_97} perturbative expansion of
\citet{Penev_Barranco_Sasselov_08a}, applied to the steady
state flow without forcing, and we found that this method also predicts linear
scaling of the effective viscosity with period, having a similar slope to the
one predicted by the direct calculation. 

Again there is a significant constant
offset between the perturbatively calculated turbulent viscosity and the
viscosity from the direct calculation. Some part of this offset might be due to
the fact that the perturbative calculation is only valid when the external
perturbation period is much smaller than the convective turnover time, and the
smallest value we are able to simulate is $T\approx0.3\tau_c$. 

At least part of this difference is due to actual amplitude dependence of the
effective viscosity, which can be seen in figure \ref{fig: A dependence} for
$K_{1313}^0(\tau_c/2)$. This amplitude dependence is possibly explained by the
fact that the external shear caused by the introduced forcing is an additional
driver of turbulence, and hence dissipation. While our calculations show clear
evidence of this amplitude dependence, it is relatively less important than the
period dependence at least in the range of amplitudes accessible to our
numerical model.

The idea that the external shear is acting as an additional driver of turbulence
suggest an explanation for why we might expect some extra viscosity at the
location of the negative side lobes of the energy transfer curves. These regions
correspond to the locations where the shear of the externally forced velocity
field is the largest. Thus, if the external forcing is driving its own
turbulence and causing its own dissipation, we would expect that extra viscosity
to appear exactly at those locations. However, this explanation would suggest
that the effect should decrease as the amplitude of the external forcing
decreases and we see that this is not the case for the fixed period variable
amplitude simulations, where the difference between the two methods seems more
or less independent of the forcing amplitude.

To summarize, for forcing periods comparable to the convective turnover time
($\tau_c$) the effective viscosity scales linearly with period and for our
convective box its $K_{1313}$ and $K_{1212}$ components can be approximated by:
\begin{eqnarray}
	K_{1313}(T)&\approx&\rho
	\left<v_z^2\right>^{1/2}H_p\left[0.23 \min\left( \frac{T}{2.3\tau_c},
	1\right) + \delta_{xz}(A)\right]
	\label{eq: nu_xz_final},\\
	K_{1212}(T)&\approx&\rho
	\left<\frac{v_x^2+v_y^2}{2}\right>^{1/2} H_p\left[ 0.23\min\left( 
		\frac{T}{2.3\tau_c}, 1\right) + \delta_{xy}(A)\right]
	\label{eq: nu_xy_final},
\end{eqnarray}
where $T$ is the period of the external forcing, and $\delta_{xz}(A)$ and
$\delta_{xy}(A)$ are constant amplitude dependent offsets that are also
dependent on the particular method for deriving effective viscosities:
\begin{center}
\begin{tabular}{c|cc|cc}
		& \multicolumn{2}{c|}{\textbf{Strong Forcing}} & 
		\multicolumn{2}{c}{\textbf{Weak Forcing}}\\
		& fitted &  dissipation matching 
		& fitted &  dissipation matching \\
\hline
$\delta_{xz}$ 	& 0.10 & 0.06 & 0.08 & 0.03 \\
$\delta_{xy}$ 	& 0.09 & 0.04 & 0.02 & -0.02\\
\end{tabular}
\end{center}

The limited spatial resolution of our box does not allow us to reliably simulate
the case of $T\ll\tau_c$. In that regime, the assumptions for Kolmogorov
turbulence hold and the effective viscosity should scale quadratically with
period \citep{Goodman_Oh_97}.

We compare the above directly calculated values to a perturbative estimate
of the effective viscosity, which also predicts linear period
dependence with the same slope, but with $\delta=0$ for all viscosity
components. 

The viscosity of equations \ref{eq: nu_xz_final} and \ref{eq: nu_xy_final} is
closer to the
\citet{Zahn_66, Zahn_89} prescription including the saturation period. However,
the fact that we see the dissipation disappearing sharply at short periods
suggests that the Zahn picture in which the dissipation is dominated by the
largest eddies present is not the correct one. If this were the case the linear
scaling seen at long periods should continue to hold for arbitrarily high
frequencies. Instead our results suggest that it is the eddies with turnover
times cloes to the external forcing period that cause most of the dissipation,
and the observed linear scaling is due to the shallower than Kolmogorov time
spectrum of the convective velocities. As a result, we expect the linear loss of
efficiency to apply only in a
limited range and for shorter periods faster loss of efficiency should occur.
This might be the resolution of the apparent discrepancy
between the dissipation necessary to explain tidal circularization and the red
edge of the Cepheid instability strip on one hand and the observed amplitudes of
the solar $p$-modes on the other. For small periods (of order minutes) the
Kolmogorov scaling of turbulence holds and a quadratic decrease in the
effective viscosity is appropriate. For long periods (of order days) the
assumptions necessary for Kolmogorov cascade are not satisfied and the effective
viscosity is found to scale linearly with period.

\appendix
\include{AppendixRuns}

\end{document}